\documentclass[11pt]{article}
\pdfoutput = 1

\usepackage[utf8]{inputenc}
\usepackage{color,graphicx}
\usepackage{epsfig}
\usepackage{amsmath}
\usepackage{verbatim}
\usepackage{amssymb}
\usepackage{mathabx}
\usepackage[mathcal]{eucal}
\usepackage{stmaryrd}
\usepackage{empheq}
\usepackage{esint}
\usepackage{physics}
\usepackage{amsfonts}
\usepackage{array}
\usepackage{setspace}
\usepackage{braket}
\usepackage{float}
\usepackage{url}
\usepackage{mathtools}
\usepackage{tensor}
\usepackage{bbold}
\usepackage{slashed}
\usepackage{tikz}
\usepackage{graphicx}
\usepackage{epstopdf}
\usepackage{subfigure}
\usepackage[labelfont=bf,font={sf}]{caption}
\usepackage{pgfplots}
\usepackage{mathrsfs}
\usepackage{fancybox}
\usepackage{eurosym}
\usepackage{tcolorbox}
\usepackage{tensor}
\allowdisplaybreaks

\usepackage[margin = 2.2cm]{geometry}
\usepackage[ragged]{footmisc}
\usepackage[bookmarks=true,bookmarksnumbered=false,
hyperindex=true,bookmarksopen=true,hyperfigures=true,
colorlinks=true,linkcolor=webblue,citecolor=webgreen,urlcolor=webblue,breaklinks]{hyperref}
\definecolor{webgreen}{rgb}{0, 0.5, 0}
\definecolor{webblue}{rgb}{0, 0, 0.5}
\definecolor{webred}{rgb}{0.5, 0, 0}
\definecolor{darkgreen}{rgb}{0,0.5,0}

\newcommand{\average}[1]{\left\langle #1 \right\rangle}

\newcommand{\dpi}{\mathcal{D}}

\def\ben{\begin{equation}}
\def\een{\end{equation}}

\let\a=\alpha  \let\g=\gamma \let\d=\delta 
   
     \let\r=v
 \let\t=\tau

\def\be{\begin{equation}}
\def\ee{\end{equation}}
\def\ba{\begin{array}}
\def\ea{\end{array}}

\def\dalemb#1#2{{\vbox{\hrule height .#2pt
       \hbox{\vrule width.#2pt height#1pt \kern#1pt
               \vrule width.#2pt}
       \hrule height.#2pt}}}

\newcommand{\bea}{\begin{eqnarray}}
\newcommand{\eea}{\end{eqnarray}}

\renewcommand{\d}{\mathrm{d}}
\renewcommand{\i}{\mathrm{i}}

\numberwithin{equation}{section}

\begin{document}

\thispagestyle{empty}
\begin{center}
    ~\vspace{5mm}

     {\LARGE \bf 
    
   The dilaton gravity hologram of double-scaled SYK
   }
    
   \vspace{0.4in}
    
    {\bf Andreas Blommaert$^1$, Thomas G. Mertens$^2$, Jacopo Papalini$^2$}

    \vspace{0.4in}
    {$^1$SISSA and INFN, Via Bonomea 265, 34127 Trieste, Italy\\
    $^2$Department of Physics and Astronomy\\
Ghent University, Krijgslaan, 281-S9, 9000 Gent, Belgium\\
    }
    \vspace{0.1in}
    
    {\tt ablommae@sissa.it, thomas.mertens@ugent.be, 	jacopo.papalini@ugent.be}
\end{center}

\vspace{0.4in}

\begin{abstract}
\noindent We work out a precise holographic duality between sine dilaton gravity, and DSSYK. More precisely, canonical quantization of sine dilaton gravity reproduces q-Schwarzian quantum mechanics, which is the auxiliary system that arises from the chord diagrams of DSSYK. The role of the chord number in DSSYK is played by the (Weyl rescaled) geodesic length in the bulk. The most puzzling aspect of reconciling DSSYK with a simple gravitational dual at the classical level is the distinction between temperature and ``fake temperature''. At the q-Schwarzian level, we clarify how this arises from the constraint that the chord number is positive. The on-shell q-Schwarzian action with the constraint reproduces the thermodynamics of DSSYK. Semi-classically, in sine dilaton gravity this translates to the insertion of a defect, from which we deduce that fake temperature is the Hawking temperature of a smooth Lorentzian black hole. We comment on several relations with dS space. One remarkable feature is that in sine dilaton gravity quantization discretizes spacetime, therefore the Hilbert space is discrete.
\end{abstract}

\pagebreak
\setcounter{page}{1}
\tableofcontents

\section{Introduction}\label{sect:intro}
The SYK model consists of $N$ Majorana fermions $\psi_i$ with the following $p$-local Hamiltonian\footnote{Following \cite{Blommaert:2023opb,Blommaert:2023wad} our convention for $q$ is slightly different from that used for instance in \cite{Lin:2022rbf} by $q^2=q_\text{there}$.}
\begin{equation}
    H_\text{SYK}=\i^{p/2}\sum_{i_1<\dots <i_p}J_{i_1\dots i_p}\psi_{i_1}\dots \psi_{i_p}\,,\quad \abs{\log q}=\frac{p^2}{N}\,,\quad 0<q<1\,.\label{1.1}
\end{equation}
The couplings $J_{i_1\dots i_p}$ are (usually) taken to be Gaussian random. This model thanks most of its fame to the fact that for fixed $p$, the low-energy physics is governed by Schwarzian quantum mechanics \cite{kitaev2015simple,Maldacena:2016hyu}; which in turn is holographically dual to JT gravity \cite{Maldacena:2016upp,Engelsoy:2016xyb,Jensen:2016pah}, recently reviewed in \cite{Mertens:2022irh}. This duality is likely the simplest realization of AdS/CFT, and has taught us a lot about black holes in (lower-dimensional) AdS. Interestingly, in a double-scaling limit where $p\to\infty$ and $N\to\infty$ whilst $0<q<1$ remains finite, SYK remains exactly solvable \cite{Cotler:2016fpe,Berkooz:2018jqr,Berkooz:2018qkz} in terms of a (relatively) simple auxiliary quantum mechanical system. This regime is called double-scaled SYK (or DSSYK), and the auxiliary quantum mechanical model \cite{Berkooz:2018jqr,Berkooz:2018qkz,Lin:2022rbf} is a version of the ``q-Schwarzian'' quantum mechanics introduced in \cite{Blommaert:2023opb,Blommaert:2023wad}. For other recent work involving DSSYK, see for instance \cite{Jafferis:2022wez,Susskind:2022bia,Bhattacharjee:2022ave,Susskind:2023hnj,Mukhametzhanov:2023tcg,Berkooz:2023cqc,Okuyama:2023bch,Lin:2022nss,Berkooz:2022mfk,Goel:2023svz,Narovlansky:2023lfz,Verlinde:2024zrh,Berkooz:2024evs,Lin:2023trc,Verlinde:2024znh,Almheiri:2024ayc,Almheiri:2024xtw}. We briefly review DSSYK and the q-Schwarzian theory in \textbf{section \ref{sect:background}}.

DSSYK or q-Schwarzian QM should be viewed as an all-energies generalization of Schwarzian QM. By all energies, we mean that DSSYK is UV-complete in the sense that its spectrum truncates. Indeed, as reviewed and discussed in \textbf{section \ref{3.1}} for the semi-classical regime $\abs{\log q}\ll 1$, the entropy of DSSYK is a non-monotonic function of energy
\begin{equation}
    S=\frac{\pi \theta-\theta^2}{\abs{\log q}}\,, \qquad
    \begin{tikzpicture}[baseline={([yshift=-.5ex]current bounding box.center)}, scale=0.7]
 \pgftext{\includegraphics[scale=1]{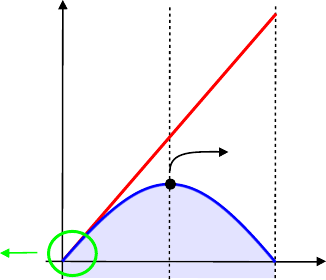}} at (0,0);
    \draw (3.7,-0.2) node {$\beta=0$ max entropy};
    \draw (2.6,-2.8) node {$\theta$};
    \draw (1.9,-2.8) node {$\pi$};
    \draw (0.1,-2.8) node {$\pi/2$};
    \draw (-1.7,-2.8) node {$0$};
    \draw (1.55,-1) node {\color{blue}$S$};
    \draw (0.8,1.8) node {\color{red}$S_\text{BH}$};
    \draw (-4.3,-1.9) node {\color{green}JT gravity};
  \end{tikzpicture} \label{1.0 entropy}
\end{equation}
The parameter $\theta$ is related to the (inverse) temperature as $\beta=\frac{2\pi-4\theta}{\sin(\theta)}$, and the energy as $E=-\frac{\cos(\theta)}{2\abs{\log q}}$.
At low energies, DSSYK reduces to Schwarzian QM. An immediate question that arises is whether the full DSSYK model (or the q-Schwarzian) also has a tractable bulk gravitational dual, which should be considered an all-energies generalization of JT gravity. In this work, we find such a gravitational dual. 

We will establish a precise holographic duality between DSSYK and 2d sine dilaton gravity:
\begin{equation}
\int \dpi g\dpi\Phi\,\exp\bigg( \frac{1}{2}\int \d x \sqrt{g}\bigg(\Phi R+\frac{\sin(2\abs{\log q} \Phi)}{\abs{\log q}}\bigg)+\text{boundary}\bigg)\, .\label{1.1 sinedil}
\end{equation}
Here $\Phi$ is the dilaton field, $R$ is the Ricci scalar of the metric $g$, and the boundary term is specified in \eqref{b+b}. We have also set $8\pi G_N =1$. The dictionary involves an effective length variable $L$. It represents the worldline action of bulk matter particles that couple non-minimally to sine dilaton gravity. In particular, we are led to consider massive particles with the following worldline action
\begin{equation}
   e^{-\Delta L}\,,\quad L=\int\d s\,e^{-\i\abs{\log q}\Phi}\,.\label{1.2 matterac}
\end{equation}
We will show that canonical quantization of sine dilaton gravity exactly reproduces DSSYK. In particular, we find that operator insertions $\mathcal{O}_\Delta$ in DSSYK correspond with inserting non-minimally coupled massive particles with action \eqref{1.2 matterac} on the holographic boundary \eqref{2.3 bc} in sine dilaton gravity \eqref{1.1 sinedil}. Our Weyl-rescaled length $L$ is identified with the ``chord number'' n \cite{Berkooz:2018jqr,Berkooz:2018qkz,Lin:2022rbf} in DSSYK as\footnote{We will use bold symbols $\mathbf{x}$ to denote quantum operators associated with classical variables $x$.} 
\begin{equation}
\mathbf{L}=2\abs{\log q} \mathbf{n}\,.\label{1.5 dictionary}
\end{equation}
Canonical quantization of sine dilaton gravity is discussed in \textbf{section \ref{sect:gravdual}}.

The entropy profile \eqref{1.0 entropy} is highly non-standard from a gravitational point of view, especially if one imagines a simple dual consisting of only black holes (as we have). Indeed, the black hole Bekenstein-Hawking entropy \cite{hawking1975particle}
\begin{equation}
    S_\text{BH}=\frac{A}{4 G_{\text{N}}}\,,
    \label{1.6 fake entropy}
\end{equation}
generically grows \emph{monotonically} with the energy $E$ of black holes. In DSSYK, the entropy $S$ does not grow monotonically with $E$ \eqref{1.0 entropy}, and differs from \eqref{1.6 fake entropy}. How is this reconciled with black hole physics? 

Aside from this, arguably the most puzzling aspect of DSSYK from a gravitational point of view is the distinction between temperature $\beta^{-1}$ and ``fake temperature'' $\beta_\text{BH}^{-1}$ \cite{Lin:2023trc} (or ``tomperature'' \cite{Lin:2022nss}):
\begin{equation}
    \beta_\text{BH}=\frac{2\pi}{\sin(\theta)}\neq \beta\,. \label{1.7 fakebeta}
\end{equation}
Temperature is (inverse) periodicity in Euclidean time. Fake (inverse) temperature is the decay time of correlators. In DSSYK (see \textbf{section \ref{3.2}}), they are different. We will show that both this distinction between temperature and fake temperature and the non-monotonic entropy follows from a constraint that the number of chords is positive. In gravitational variables, our length operator is constrained to be non-negative
\begin{equation}
    \boxed{\mathbf{L}\geq 0}\,.\label{1.7 constraint}
\end{equation}
This is a non-standard restriction from a gravitational point of view. In particular, we show in \textbf{section \ref{2}} that \emph{without} imposing this positivity condition, the entropy of the theory would grow monotonically and equal $A/4 G_N$; and that temperature, fake temperature and Bekenstein-Hawking temperature coincide!

A more semi-classical bulk gravitational interpretation of this positivity condition is desirable. In particular, we want to pinpoint how at the semi-classical level black hole physics produces a distinction between temperature and fake temperature, and relatedly how the bulk can be reconciled with DSSYK being a sub-maximal scrambler \cite{Maldacena:2016hyu,Streicher:2019wek,Choi:2019bmd}. We show in \textbf{section \ref{sect:defect}} that the positivity constraint is implemented at the \emph{semi-classical} level in sine dilaton gravity by the insertion of a conical defect with opening angle $\gamma$ equal to
\begin{equation}
    \gamma=2\pi-4\theta\,.
\end{equation}
A key point \cite{Dong:2022ilf} is that, even though the defects are singular sources, the Lorentzian spacetime probed by matter is a smooth black hole with Hawking temperature equal to the fake temperature. Therefore, all matter correlators are to be expected to be thermal at the \emph{fake} temperature. Even though our bulk geometry is not AdS$_2$, this semi-classical picture for the bulk involving a defect is close in spirit to the ``fake disk'' \cite{Lin:2023trc}.

Our sine dilaton gravity model \eqref{1.1 sinedil} is related with a recent proposal by H. Verlinde \cite{Verlinde:2024znh} involving dS$_3$. In dS gravity, the distinction between fake temperature and temperature seems to be intimately related with the fact that in cosmology the observer should be included in the quantum description of the system \cite{Chandrasekaran:2022cip,Witten:2023xze}. As mentioned, the difference between $\beta$ and $\beta_\text{BH}$ follows entirely from the constraint $\mathbf{L}\geq 0$ \eqref{1.7 constraint}. So, this suggests that the constraint implements somehow the inclusion of an observer in dS space (if DSSYK is a theory of dS space \cite{Susskind:2021esx,Susskind:2022bia,Lin:2022nss,Rahman:2022jsf,Susskind:2022dfz,Verlinde:2024znh,Narovlansky:2023lfz,Verlinde:2024zrh}). We sketch how this might be realized in our model,\footnote{It would independently be interesting to understand if something along these lines is realized in \cite{Verlinde:2024znh} which also implicitly implements such a positivity constraint.} and comment further on relations with dS in the discussion \textbf{section \ref{sect:discnew}}.

One result of our analysis in \textbf{section \ref{sect:harlowjafferissine}} is that the Hilbert space of sine dilaton gravity is discrete. This is obvious from the DSSYK perspective, but is a non-trivial property for any gravitational theory. Indeed we find that at the quantum level, the geometric variable $L$ is discretized.

Before summarizing the structure of our work in \textbf{section \ref{sect:sum}}, we present some background material on the exact solution of DSSYK \cite{Berkooz:2018jqr,Berkooz:2018qkz}, and its relation with the q-Schwarzian theory \cite{Blommaert:2023opb,Blommaert:2023wad}.

\subsection{Background material}\label{sect:background}
As shown in \cite{Berkooz:2018jqr,Berkooz:2018qkz} and summarized pedagogically in \cite{Lin:2022rbf,Verlinde:2024znh}, the partition function and all correlation functions of DSSYK are most efficiently computed using an auxiliary quantum mechanics system. This quantum mechanical system is a tool to compute tedious summations of Feynman diagrams which arise in the ``chord diagram'' evaluation of correlation functions. In particular, the partition function of DSSYK is computed in the (auxiliary) quantum mechanical model as a ``transition matrix element''
\begin{equation}
    \text{Tr}\big(e^{-\beta H_\text{SYK}}\big)=\bra{\mathbf{n}=0}e^{-\beta \mathbf{H}}\ket{\mathbf{n}=0}\,,
\end{equation}
with $[\mathbf{n},\mathbf{p}]=\i$ and Hamiltonian
\begin{equation}
    \mathbf{H}=-\frac{\cos(\mathbf{p})}{2\abs{\log q}}+\frac{1}{4\abs{\log q}}e^{i\mathbf{p}}e^{-2\abs{\log q}\mathbf{n}}\,.\label{1.10 hamil}
\end{equation}
The two-point correlation function of operators $\mathcal{O}_\Delta$ is computed in this quantum mechanical model as\footnote{The interesting operators $\mathcal{O}_\Delta$ have $0<\Delta<1$ and involve (Gaussian) random numbers $M_{i_1\dots i_{\Delta p}}$
\begin{equation}
    \mathcal{O}_\Delta=\i^{\Delta p/2}\sum_{i_1<\dots <i_{\Delta p}}M_{i_1\dots i_{\Delta p}}\psi_{i_1}\dots \psi_{i_{\Delta p}}\,.
\end{equation}
}
\begin{equation}
    \text{Tr}\big(e^{-\tau H_\text{SYK}}\mathcal{O}_\Delta e^{-(\beta-\tau) H_\text{SYK}}\mathcal{O}_\Delta\big)=\bra{\mathbf{n}=0}e^{-\tau \mathbf{H}}\,e^{-2\Delta \abs{\log q} \mathbf{n}}\,e^{-(\beta-\tau)\mathbf{H}} \ket{\mathbf{n}=0}\,.\label{1.12 twopoint}
\end{equation}
The Hamiltonian $\mathbf{H}$ of the system is oftentimes called ``transfer matrix''. Readers may recognize \eqref{1.10 hamil} better from other literature by writing it using normalized raising and lowering operators $\mathbf{a}_\text{norm}=-e^{\i\mathbf{p}}$ such that $\mathbf{a}_\text{norm}\ket{n}=\ket{n-1}$. Indeed this brings \eqref{1.10 hamil} into the form
\begin{equation}
\label{eq:chordham}
    \abs{\log q}\,\mathbf{H}=\mathbf{a}_\text{norm}^\dagger + \mathbf{a}_\text{norm}\,(1-q^{2\mathbf{n}})\,.
\end{equation}
This is equation (11) for the DSSYK transfer matrix in \cite{Lin:2022rbf}.\footnote{The two expressions agree up to a harmless similarity transformation which preserves the identity
\begin{equation}
    \mathbb{1}=\sum_{n=0}^\infty \ket{n}\bra{n}\,.
\end{equation}
These does not change Hamiltonian (classical or quantum) mechanics. We thank Adam Levine for raising this point. 
} 
The variable $n$ is the number of ``chords'' present in the chord diagram at a given time slice. To clarify this, in the chord diagrams picture \cite{Berkooz:2018jqr,Berkooz:2018qkz}, for instance the two-point correlator is computed by summing over chord (Feynman) diagrams such as
\begin{align}
    \begin{tikzpicture}[baseline={([yshift=-.5ex]current bounding box.center)}, scale=0.7]
 \pgftext{\includegraphics[scale=1]{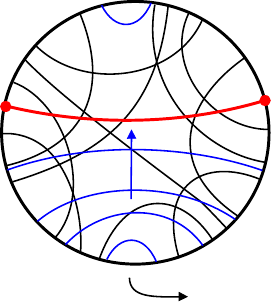}} at (0,0);
    \draw (3.4,-1.2) node {\color{blue}time flow};
    \draw (-2.9,0.5) node {\color{red}$\mathcal{O}_\Delta$};
    \draw (2.9,0.5) node {\color{red}$\mathcal{O}_\Delta$};
    \draw (3.5,-2.5) node {initial state $\ket{\mathbf{n}=0}$};
  \end{tikzpicture}\label{3.79}
\end{align}
The number of chords (black lines) intersecting a ``time slice'' (blue lines) is $n$. To us, the chords are an intermediate tool to obtain the quantum mechanical model \eqref{1.10 hamil}. We will identify the holographic dual of this quantum mechanical model. One crucial corollary of the derivation of the quantum mechanical model \eqref{eq:chordham} from chord diagrams is that the number of chords is constrained to be non-negative
\begin{equation}
    \mathbf{n}\geq 0\,,
\end{equation}
and that correlators of DSSYK map to transition matrix elements in the QM with initial and final state $\ket{\mathbf{n}=0}$. As mentioned already, these features have important implications for bulk physics.

The auxiliary quantum system \eqref{eq:chordham} is related to the q-Schwarzian theory, introduced in  \cite{Blommaert:2023opb,Blommaert:2023wad}, as follows. Introducing the variable
\begin{equation}
    \varphi=\abs{\log q} \, n\,,\label{1.19 dic}
\end{equation}
DSSYK correlators such as \eqref{1.12 twopoint} can be rewritten in terms of a path integral as
\begin{equation}
\underset{\substack{\varphi(0)=\varphi(\beta)=0}}{\int} \dpi \varphi \dpi p\dots \exp\bigg\{\frac{1}{\abs{\log q}}\int_0^\beta \d u \bigg(\i\, p \frac{\d}{\d u}\varphi+\frac{1}{2}\cos(p)-\frac{1}{4}e^{\i p}e^{-2\varphi}  \bigg) \bigg\}\,.\label{1.20 bis}
\end{equation}
Here, the $\dots$ denote operator insertions. For the two-point function \eqref{1.12 twopoint}, one inserts $e^{-2\Delta \varphi(\tau)}$. This generalizes to higher point functions with distinct $\Delta_i$ where the operator lines (see \eqref{3.79}) do not cross simply as $e^{-2\Delta_1 \varphi(\tau_1)}\dots e^{-2\Delta_n \varphi(\tau_n)}$.

Note that the paths in this path integral are not discretized. It is useful to contrast our \eqref{1.10 hamil} and the chord discretization with more conventional dynamical systems as follows.
In condensed matter physics, in the tight-binding approximation electrons are considered to be heavily bound to atoms in a periodic array.\footnote{We thank Thomas Tappeiner for a discussion on this.} Their position space wavefunctions are then approximately supported only on a discrete lattice $x = na \in \mathbb{Z}$, with $a$ the lattice spacing. This leads to the conjugate momentum $p$ being periodic $p\sim p+2\pi \hbar/a$ and one can restrict to the first Brillouin zone. For the DSSYK Hamiltonian \eqref{1.10 hamil} on the other hand, we are in the exact opposite situation: $p \sim p +2\pi$ is a classical symmetry, leading to the position space wavefunctions only having support on $x \in \hbar \mathbb{Z}$, an integer lattice with $\hbar$ spacing. In the classical $\hbar \to 0$ limit of the phase space (which is what the paths in the path integral are probing), the lattice disappears and the continuum reemerges.

An elementary but insightful derivation of \eqref{1.20 bis} as the Feynman path integral description of a canonical DSSYK evolution amplitude is presented in appendix \ref{app:fpi}.

Rescaling to the conjugate coordinate of $\varphi$ as
\begin{equation}
    p=\abs{\log q}p_\varphi\,,
\end{equation}
leads to a version of the q-Schwarzian theory presented in \cite{Blommaert:2023opb}\footnote{Unlike in \cite{Blommaert:2023opb} where time was rescaled we will use the standard time parameter of DSSYK.}
\begin{equation}
\underset{\substack{\varphi(0)=\varphi(\beta)=0}}{\int} \dpi \varphi \dpi p_\varphi\dots \exp\bigg\{\int_0^\beta \d u \bigg(\i\, p_\varphi \frac{\d}{\d u}\varphi+\frac{1}{2\abs{\log q}}\cos(\abs{\log q}p_\varphi)-\frac{1}{4\abs{\log q}}e^{\i \abs{\log q}p_\varphi}e^{-2\varphi}  \bigg) \bigg\}\,.\label{1.20 qschtheory}
\end{equation}

\subsection{Structure}\label{sect:sum}
The remainder of this work is organized as follows.

In \textbf{section \ref{2}} we investigate the duality between the q-Schwarzian and sine dilaton gravity \emph{ignoring} the positivity constraint, and show that this produces \emph{fake} thermodynamics. We also show that semi-classical correlators in the q-Schwarzian map to non-minimally coupled bulk particle worldlines \eqref{1.2 matterac}.

In \textbf{section \ref{sect:fakebdy}} we point out that semi-classical thermodynamics and correlators of DSSYK \emph{mismatch} with the sine dilaton gravity analysis of section \ref{2}, but that \emph{all} differences trace back to the distinction between temperature and fake temperature in DSSYK. This discrepancy is resolved by imposing the positivity constraint. We will show indeed that at the classical level imposing this constraint on the q-Schwarzian boundary theory reproduces DSSYK semiclassics. This is a non-trivial check on the validity of \eqref{1.20 bis}.

In \textbf{section \ref{sect:gravdual}} we work out a direct canonical quantization of sine dilaton gravity, following the quantization by D. Harlow and D. Jafferis for JT gravity \cite{Harlow:2018tqv}. A key step is to identify the Weyl-rescaled length $L$ \eqref{1.2 matterac} as a convenient coordinate on the sine dilaton gravity phase space. One finds the following simple gravitational Hamiltonian (or ADM energy E) in terms of $L$ and its conjugate $p$
\begin{equation}
   \mathbf{H}_\text{grav}=-\frac{\cos(\mathbf{P})}{2\abs{\log q}}+\frac{1}{4\abs{\log q}}e^{i\mathbf{P}}e^{-\mathbf{L}} \, .
\end{equation}
This is exactly the DSSYK auxiliary quantum mechanics Hamiltonian \eqref{1.10 hamil} with the dictionary \eqref{1.5 dictionary}! DSSYK operators $q^{2\Delta \mathbf{n}}$ map to matter particle insertions $e^{-\Delta \mathbf{L}}$ \eqref{1.2 matterac}. DSSYK computations such as \eqref{1.12 twopoint} show that we should impose the constraint $\mathbf{L}\geq 0$ in gravity, and that DSSYK observables map to gravitational transition matrix elements. We then present the classical sine dilaton gravity defect interpretation of the constraint (and transition matrix element prescription), and how this addresses the temperature vs fake temperature conundrum.

In \textbf{section \ref{sect:5.1}} we write down the relation between the sine dilaton gravity action and two copies of the 2d Liouville action, linking it to other recent work in the field.

In \textbf{section \ref{sect:discnew}} we comment on relations of our theory with dS space. We also touch on generalizations and future directions, such as identifying a dual QM for dilaton gravity models with more generic potentials, and a possible relation of sine dilaton gravity with the recently discussed quantum disk \cite{Almheiri:2024ayc}.

\section{The q-Schwarzian and sine dilaton gravity}\label{2}
We start with delving into the equivalence between sine dilaton gravity and the q-Schwarzian, \emph{ignoring} the positivity constraint \eqref{1.7 constraint}. In section \ref{2.1} and section \ref{2.2}, we find a semi-classical match of their partition functions and see the fake DSSYK thermodynamics emerge. This follows closely the analysis for $q=e^{\i\pi b^2}$ in \cite{Blommaert:2023wad}. We refer the reader to \cite{Blommaert:2023wad} for technical details, and instead will focus on highlighting new aspects that arise in our current $0<q<1$ case (relevant for DSSYK). In section \ref{sect:operatordic} we provide semi-classical evidence for the dictionary 
\begin{equation}
    L=2\varphi\,,\label{2.1 dictionary}
\end{equation}
with $\varphi$ the q-Schwarzian variable used in \eqref{1.20 qschtheory} and $L$ the effective geodesic length \eqref{1.2 matterac}. In particular, we observe that q-Schwarzian correlators match boundary-to-boundary propagators of a bulk massive particle non-minimally coupled to sine dilaton gravity. One key ingredient in finding this match is the location of the holographic screen in sine dilaton gravity, which we also discuss in section \ref{2.1}. 

We stress that we will prove the quantum equivalence of the q-Schwarzian and sine dilaton gravity in section \ref{sect:gravdual}. The purpose of this section is mainly to show that on a classical level, \emph{ignoring} the positivity constraint \eqref{1.7 constraint} removes many of the important features of DSSYK, namely the non-monotonicity of the entropy \eqref{1.0 entropy} and (relatedly) the distinction between temperature and fake temperature.

\subsection{Sine dilaton gravity semiclassics }\label{2.1}
Let us consider sine dilaton gravity \eqref{1.1 sinedil}
\begin{equation}\label{b+b}
    \int \dpi g\dpi\Phi\,\exp\bigg( \frac{1}{2}\int \d x \sqrt{g}\bigg(\Phi R+\frac{\sin(2\abs{\log q} \Phi)}{\abs{\log q}}\bigg)+\int \d\tau \sqrt{h} \bigg(\Phi K-\i\,\frac{e^{-\i \abs{\log q}\Phi}}{2\abs{\log q}}\bigg)\bigg)\,,
\end{equation}
supplemented by the boundary conditions
\begin{equation}
    \sqrt{h}\,e^{\i\abs{\log q}\Phi_\text{bdy}}=\frac{\i}{2\abs{\log q}}\,,\qquad 2\abs{\log q} \Phi_\text{bdy}= \frac{\pi}{2}+\i\infty\,.\label{2.3 bc}
\end{equation}
Compared to \eqref{1.1 sinedil}, we explicitly included the GHY boundary term and added the appropriate counterterm required for holographic renormalization. The first boundary condition is fixing the asymptotics of the bulk geometry, whereas the second condition is fixing the asymptotics of the dilaton field.

Our choice of boundary conditions \eqref{2.3 bc} and the associated counterterm in \eqref{b+b} can be motivated in several complementary ways. 
Firstly, any bulk dilaton gravity model can be written in its first-order formulation as a non-linear gauge theory, the Poisson-sigma model. Within that language, natural gravitational boundary conditions are so-called mixed parabolic boundary conditions and this is precisely our choice of boundary condition. We present these arguments in appendix \ref{s:PSM}.

Secondly, as we show in section \ref{sect:operatordic}, q-Schwarzian correlators (and DSSYK correlators, see section \ref{3.1}) have the form of free matter boundary propagators in an AdS$_2$ geometry. This can be explained by matter that couples non-minimally to sine dilaton gravity, see \eqref{1.2 matterac}. Essentially, this boils down to considering a Weyl-rescaled bulk metric, which is AdS$_2$. But even with that Weyl rescaling, the dictionary would of course only work if the operators are inserted on the asymptotic AdS$_2$ boundary. The location $\Phi_\text{bdy}$ is simply the translation of the asymptotically AdS$_2$ boundary in the Weyl rescaled metric, to the original sine dilaton metric. A non-trivial check is that this choice of boundary location \emph{implies} the boundary conditions (the first entry in \eqref{2.3 bc}), if we look at the metric solutions \eqref{metrrr}. 

A final argument for our boundary conditions is in terms of the Liouville string description, as we will discuss in section \ref{sect:5.1}.

We next discuss the classical solutions of \eqref{b+b}. Upon rescaling $ 2\abs{\log q} \Phi \rightarrow \Phi$ one finds
\begin{equation}\label{sine_path}
    \int \dpi g\dpi\Phi\,\exp\bigg( \frac{1}{2 \abs{\log q}}\left\{\frac12\int \d x \sqrt{g}\,\big(\Phi R+2\sin(\Phi)\big)+\int \d\tau \sqrt{h}\,\big(\Phi K-\i\,e^{-\i \Phi/2}\big)\right\}\bigg)\,,
\end{equation}
Because of the overall $1/\abs{\log q}$, the classical saddle is reliable for $\abs{\log q}\ll 1$, which is therefore our semi-classical regime of interest.\footnote{This does not reduce to JT gravity because we consider finite energies. JT gravity is recovered upon $\abs{\log q}\ll 1$ \emph{without} rescaling the dilaton.} The classical solutions for general dilaton gravity models are well-known \cite{Gegenberg:1994pv,Witten:2020ert}. For our specific model \eqref{sine_path} one obtains
\begin{align}
        \d s^2&=F(r)\d \tau^2+\frac{1}{F(r)}\d r^2\,,\qquad F(r)=-2 \cos(r)+2 \cos(\theta)\,,\qquad \Phi=r\,. \label{metrrr}
\end{align}
Notice that this metric satisfies indeed (the correctly rescaled version of) our boundary conditions \eqref{2.3 bc}
\begin{equation}
\sqrt{F}\,e^{\i\Phi_\text{bdy}/2}=\i\,,\quad\Phi_\text{bdy}=\frac{\pi}{2}+\i\infty\,.\label{2.6 bcrescaled}
\end{equation}

This solution \eqref{metrrr} has a black hole horizon at $r=\theta$. Expanding the line element around the horizon by defining $r=\theta+\rho^2$ with $\rho\ll 1$ one obtains
\begin{align}\label{near-horizon}
 \d s^2&=\frac{2}{\sin(\theta)} (\d \rho^2+\sin^2 (\theta)\, \rho^2 \d \tau^2 )\,.
\end{align}
By demanding that the Euclidean section of \eqref{near-horizon} has no conical deficit as we move around the thermal circle $\tau\sim \tau+\beta_{\text{BH}}$, we deduce that the Hawking temperature associated with this horizon is
\begin{equation}
    \beta_\text{BH}=\frac{2\pi}{\sin(\theta)}\,.\label{2.8 smooth}
\end{equation}
One can study the classical thermodynamics of these solutions by plugging \eqref{metrrr} into the action \eqref{sine_path}, and integrating along any contour in the complex $r$ plane from $r=\theta$ to $r=\Phi_\text{bdy}$. This leads to the following on-shell action for fixed $\theta$ and $\beta$
\begin{align}\label{onshell}
    \exp\bigg(\frac{1}{2 \abs{\log q}}\bigg\{2\pi\theta+\beta F(\Phi_\text{bdy})-\i\, \beta F(\Phi_\text{bdy})^{1/2}e^{-\i \Phi_\text{bdy}/2}\bigg\}\bigg) \,.
\end{align}
The last term in this expression is the counterterm which was introduced in \eqref{b+b}.\footnote
{
For a generic dilaton gravity model, the relevant boundary term that produces classical thermodynamics is given by
\begin{equation}
\int \d\tau \sqrt{h}\,\left(\Phi K-\sqrt{G(\Phi)}\right), \qquad G(\Phi_\text{bdy}) = \int^{\Phi_{\text{bdy}}}\d \Phi\,V(\Phi)\,.
\end{equation}
Indeed, in our case when we insert $V(\Phi) = 2 \sin \Phi$ and the asymptotics of the dilaton field \eqref{2.6 bcrescaled}, we recover precisely our choice of boundary term \eqref{sine_path}.
}
Expanding around the asymptotic value $\Phi_\text{bdy}=\pi/2+\i\infty$, and integrating over $\theta$, one finds indeed a finite expression\footnote{One-loop factors are not tracked, hence our choice of measure $\d E(\theta)$ is in principle left arbitrary.}
\begin{equation}
    Z_\text{grav}(\beta)\overset{\text{class}}{=}\int\d E(\theta)\,\exp\bigg(\frac{\pi \theta}{\abs{\log q}}+\beta \frac{\cos(\theta)}{2\abs{\log q}}\bigg)\,.\label{2.10 td}
\end{equation}
From this expression one reads off the ADM energy and the black hole area\footnote{
This also follows from the generic dilaton gravity thermodynamics formulas, as written down e.g. in section 2.3.2 in \cite{Mertens:2022irh}:
\begin{equation}
E(T) = \frac{1}{2}\int^{V^{-1}(4\pi T)} \hspace{-0.2cm}V(\Phi)\,\d\Phi = \sqrt{1-4\pi^2 \beta^{-2}}\,.
\end{equation}
from which all thermodynamic relations follow.
}
\begin{equation}
     E=-\frac{\cos(\theta)}{2\abs{\log q}}\,,\qquad S_\text{BH}=  \frac{\pi \theta}{\abs{\log q}}\,.\qquad\qquad 
  \label{2.bb energy}
\end{equation}
Notice that this ADM energy is also identified as the subleading part of the metric which remains finite near the holographic boundary (taking into account the rescaling factor $2\abs{\log q}$), generalizing the usual notion for asymptotically AdS spacetimes \cite{brown1993quasilocal}, as
\begin{equation}
    F(\Phi_\text{bdy})=-e^{-\i\Phi_\text{bdy}}+4\abs{\log q} E\,.
\end{equation}
These thermodynamical relations \eqref{2.bb energy} reproduce \emph{fake} DSSYK thermodynamics \eqref{1.6 fake entropy} and \eqref{1.7 fakebeta}  (which we will discuss in detail in section \ref{3.1}).\footnote{This difference between fake and true thermodynamics can not be cured by adding a different counterterm that subtracts the divergent part in \eqref{onshell}, there is only an ambiguity in a constant shift of ADM energy.}

Let us briefly stress some interesting features of the classical solutions \eqref{metrrr}. Notice that this metric has a black hole horizon at $r=\theta$ \emph{and} a cosmological horizon at $r=2 \pi-\theta$ (with respectively minimal and maximal area). In fact, there is an infinite set of horizons on the $r$ axis, obtained from the ``original'' ones by shifting $\theta$ by $2\pi n$. The spacetime with the integration contour from $r=\theta$ to $r=\Phi_\text{bdy}$ hence looks as follows:
\begin{equation}
   \begin{tikzpicture}[baseline={([yshift=-.5ex]current bounding box.center)}, scale=0.7]
 \pgftext{\includegraphics[scale=1]{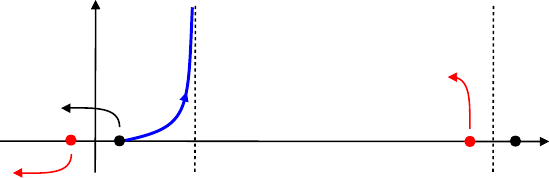}} at (0,0);
    \draw (1.2,-0.15) node {\color{red}horizon};
    \draw (1.2,0.55) node {\color{red}cosmological};
    \draw (-4.8,-0.25) node {horizon};
    \draw (-5.4,-1.45) node {\color{red}more};
    \draw (-5.4,-2.15) node {\color{red}horizons};
    \draw (4.6,-1.8) node {$r$};
    \draw (3.7,-1.8) node {$2\pi$};
    \draw (-1.3,-1.8) node {$\pi/2$};
    \draw (-3.05,-1.8) node {$0$};
    \draw (-1.3,2.7) node {\color{blue}integration};
    \draw (-1.3,2) node {\color{blue}contour};
  \end{tikzpicture} \label{2.bb horizons}
\end{equation}
Here we drew an arbitrary complex contour that connects the horizon with the (complex) holographic boundary. As we tend to observe real geometries, it makes sense to align the contour with real metrics ``as much as possible''. We will continue this discussion in \ref{sect:5.2}. Before moving on, we note that according to the work of \cite{Kruthoff:2024gxc} one is supposed to sum over horizons in dilaton gravity. So, in our case one should include the saddle where the contour in the $r$ plane starts at the cosmological horizon $r=2\pi-\theta$ and ends at $\Phi_\text{bdy}$. Should we also count spacetimes that end on the infinite set of horizon copies? The exact DSSYK spectral density \eqref{4.34 magic} suggests so. We leave a better understanding of the role of this infinite series of horizons to future work.

\subsection{q-Schwarzian semiclassics }\label{2.2}
Next, we want to study a similar semi-classical limit of the q-Schwarzian theory \eqref{1.20 bis}, again \emph{ignoring} the positivity constraint \eqref{1.7 constraint}, and compare the results with the sine dilaton gravity semiclassics studied in this section. This subsection parallels our earlier work on Liouville gravity in \cite{Blommaert:2023wad}.

To avoid awkward factor of $\i$, let us consider \eqref{1.20 bis} with $k=\i p$ and without putting any constraints on $\varphi$
\begin{equation}
    \int \dpi \varphi \dpi p\, \exp\bigg\{\frac{1}{\abs{\log q}}\int_0^\beta \d \tau \bigg(k \frac{\d}{\d \tau}\varphi+\frac{1}{2}\cosh(k)-\frac{1}{4}e^{k}e^{-2\varphi}  \bigg) \bigg\}\,.\label{2.19 bis}
\end{equation}
This (two-sided) theory was referred to as q-Liouville in \cite{Blommaert:2023opb}, and just like the gravitational model \eqref{sine_path} it should have a semi-classical regime dominated by saddles for $\abs{\log q}\ll 1$. The equations of motion can be massaged into the simple form
\begin{equation}
    \frac{\d^2}{\d \tau ^2}\varphi=-\frac{1}{4}e^{-2\varphi}\,,\quad \frac{\d}{\d\tau}e^{-k}=-\frac{1}{2}e^{-2\varphi}\,.
\end{equation}
The first equation is Liouville's equations whose solutions are AdS$_2$ conformal factors. Introducing an integration constant $\theta$ one finds (see also \cite{Blommaert:2023wad}):
\begin{align}\label{sol}
    e^{-2\varphi}=\frac{\sin(\theta)^2}{\sin(\sin(\theta)\tau/2)^2}\,,\quad e^{-k}=\frac{\sin(\theta)}{\tan(\sin(\theta)\tau/2)}+\cos(\theta)\,,\quad 0<\tau<\frac{2\pi}{\sin(\theta)}\,.
\end{align}
In addition, one finds different solutions by shifting $\tau$ by a constant. At this point, we are not interested in them, but they are relevant in section \ref{sect:3.3 new}.\footnote{Here, one should think of this constant as fixed by demanding that the solutions \eqref{sol} satisfy $\varphi(0)=\varphi(\beta)=-\infty$.} 
The period of the solutions depends on the integration constant $\theta$. One checks explicitly that on these solutions the Hamiltonian is a constant
\begin{equation}
    E=-\frac{\cosh(k)}{2\abs{\log q}}+\frac{1}{4\abs{\log q}}e^{k}e^{-2\varphi}=-\frac{\cos(\theta)}{2\abs{\log q}}\,.\label{2.22 en}
\end{equation}
We already notice that this agrees with the ADM energy of the sine dilaton theory in \eqref{2.bb energy}, and that the period of the solution matches the Hawking temperature of the black hole in sine dilaton gravity \eqref{2.8 smooth}. The remainder of the on-shell action should match the entropy \eqref{2.bb energy} of sine dilaton gravity. We want to compute the entropy $S_\text{qsch}$ through
\begin{equation}\label{entropy}
    e^{S_\text{qsch}}=\exp\bigg(\frac{1}{\abs{\log q}}\oint k \d\varphi \bigg) \, .
\end{equation}
It is convenient to evaluate this integral over the closed orbit using complex analysis. Introducing the complex variable
\begin{equation}
    z=e^{\i\sin(\theta)\tau}\, ,
\end{equation}
one rewrites
\begin{equation}
    \frac{\d}{\d z}\varphi=-\frac{1}{2 z}-\frac{1}{1-z}\,,\quad k=-\log \bigg(-\i\sin(\theta)\frac{1+z}{1-z}+\cos(\theta)\bigg)\,.
\end{equation}
The branch cut of $k$ goes from $z=1$ to $z=e^{-2\i \theta}$. Then the entropy \eqref{entropy} is written as
\begin{equation}\label{f(z)}
S_{\mathrm{qsch}}=\frac{1}{2\abs{\log q}}\oint_\gamma \d z\, \bigg(\frac{1}{z}+\frac{2}{1-z}\bigg) \log \bigg(-\i\sin(\theta)\frac{1+z}{1-z}+\cos(\theta)\bigg)\, .
\end{equation}
We claim that the relevant contour $\gamma$ should be taken to run to the left of the branch cut of $k$, such that this integral only picks up the pole at $z=0$ (which corresponds to $T=+\infty$ in the classical solutions). Using
\begin{equation}
    \oint_0 \frac{\d z}{z} \log \bigg(-\i\sin(\theta)\frac{1+z}{1-z}+\cos(\theta)\bigg)=2\pi\theta=2\pi p(z=0)\,, 
\end{equation}
one indeed recovers the black hole entropy computed in \eqref{2.bb energy}. The q-Schwarzian partition function in the semi-classical approximation is thus given by
\begin{equation}
Z_\text{qsch}(\beta)\overset{\text{class}}{=}\int\d E(\theta)\,\exp\bigg(\frac{\pi \theta}{\abs{\log q}}+\beta \frac{\cos(\theta)}{2\abs{\log q}}\bigg)\,,\label{2.28 zqschnaive}
\end{equation}
which indeed matches precisely with the sine dilaton gravity answer \eqref{2.10 td}.
\subsection{Holographic dictionary from semi-classical correlators }\label{sect:operatordic}
After analyzing the partition function, we move on in constructing the holographic dictionary between sine dilaton gravity and the q-Schwarzian. Our focus is now on semi-classical correlators, which suggest the identification $L=2\varphi$ \eqref{2.1 dictionary}, as we will now demonstrate.

As mentioned in section \ref{sect:background}, DSSYK correlators corresponds to the insertion of the operator $e^{-2 \Delta \phi(\tau)}$ in the q-Schwarzian path integral \eqref{1.20 bis}. So, at the semi-classical level, the computation of correlators boils down to evaluating these operators on the saddle of the path integral for $\abs{\log q}\rightarrow 0$. This saddle is valid under the assumption that $\Delta$ is finite for $\abs{\log q}\rightarrow 0$, such that the operator insertions do not backreact. Hence, according to \eqref{sol} the semi-classical q-Schwarzian correlators are products of 
\begin{align}\label{ads}
  e^{-2 \Delta\varphi}=\frac{\sin(\theta)^{2\Delta}}{\sin(\sin(\theta)\tau/2)^{2\Delta}}.
\end{align}
We notice that this is on the nose the boundary-to-boundary propagator of a massive particle in AdS$_2$, with temperature
\begin{equation}
    \tau\sim \tau +\frac{2\pi}{\sin(\theta)}=\tau+\beta_{\mathrm{BH}}.
\end{equation}
While the correlator looks thermal at the temperature associated with the black hole solution \eqref{metrrr} for sine dilaton gravity, the AdS$_2$ structure requires further explanation. Specifically, from the holographic perspective, we want to know the type of bulk probe that corresponds to this quantity in sine dilaton gravity. The answer is a massive particle non-minimally coupled to the metric, meaning it also couples to the dilaton field.\footnote{We thank Ohad Mamroud for discussions about this.} In particular, we consider a scalar matter field that couples as
\begin{equation}\label{non_minimal}
    \int \d^2 x \sqrt{g}\,\bigg(g^{\mu\nu}\partial_\mu \phi \partial_\nu \phi +m^2 e^{-2\i \abs{\log q}\Phi}\phi^2\bigg)= \int \d^2 x \sqrt{g_\text{eff}}\,\bigg(g^{\mu\nu}_\text{eff}\partial_\mu \phi \partial_\nu \phi +m^2 \phi^2\bigg)\,,
\end{equation}
where in the second step we absorbed the coupling to the dilaton into an \emph{effective} metric that the probe experiences
\begin{equation}
    \d s^2_\text{eff}=e^{-2\i \abs{\log q}\Phi}\,\d s^2 \,.
\end{equation}
In the worldline particle formalism, the boundary-to-boundary propagator would be computed by the path integral over boundary-anchored paths weighted by the single particle action
\begin{equation}\label{particle_action}
   \int \dpi x\, \exp\bigg(-m \int_{x(s)} \d s\,e^{-\i \abs{\log q}\Phi}\bigg).
\end{equation}
The classical saddles of this action are geodesics in the \emph{effective} geometry $\d s^2_\text{eff}$, the length of which we denote by $L$
\begin{equation}
    \boxed{L=\int\d s\,e^{-\i\abs{\log q}\Phi}}\,.
\end{equation}

The point is that $\d s^2_\text{eff}$ is an AdS$_2$ disk. Indeed, we recall that after rescaling $2\abs{\log q}\Phi\to \Phi$ the sine dilaton gravity solutions are \eqref{metrrr}
\begin{equation}
    \d s^2=F(r)\d \t^2+\frac{1}{F(r)}\d r^2\,,\quad F(r)=2\cos(\theta)-2\cos(r)\,.\label{2.35 metr}
\end{equation}
Introducing
\begin{equation}
    r=\frac{\pi}{2}+\i \log (\rho+\i\cos(\theta))\,,
\end{equation}
the effective geometry $\d s^2_\text{eff}=e^{-\i \Phi}\d s^2$ takes the form of an AdS$_2$ black hole with Hawking temperature $\beta_\text{BH}=2\pi/\sin(\theta)$:\footnote{An alternative equivalent way to appreciate this is to use the Liouville string language of section \ref{sect:5.1}, the effective metric is just the metric corresponding to one of the two Liouville fields, with a classical AdS$_2$ solution indeed:
\begin{equation}
\d s^2_{\text{eff}} = e^{2ib \chi} \d z \d \bar{z}\, .
\end{equation}
}
\begin{equation}\label{eff}
    \d s^2_\text{eff}=F_\text{AdS}(\rho)\d\t^2+\frac{1}{F_\text{AdS}(\rho)}\d \rho^2\,,\quad F_\text{AdS}(\rho)=\rho^2-\sin(\theta)^2\,.
\end{equation}
This spacetime has a black hole horizon at $\rho=\sin(\theta)$. In addition, as usually, there is an ``interior''\footnote{This was called inner horizon in \cite{Kruthoff:2024gxc}.} horizon at $\rho=-\sin(\theta)$. The black hole horizon in the effective geometry maps to the black hole horizon $r=\theta$ in the sine dilaton metric \eqref{2.35 metr}. The interior horizon $\rho=-\sin(\theta)$ stems from an interior horizon at $r=-\theta$, which one could think of as a duplicate of the cosmological horizon at $r=2\pi-\theta$. Finally, the boundary location \eqref{2.6 bcrescaled} $r_\text{bdy}=\pi/2+\i\infty$ simply maps to the usual asymptotically AdS$_2$ boundary
\begin{equation}
    \rho_\text{bdy}=\infty\,.
\end{equation}
In summary, the standard real $\rho$ AdS$_2$ contour maps to the following complex $r$ contour in sine dilaton gravity (this is homologous to the contour discussed in section \ref{2.1}, because the boundary points agree)
\begin{equation}
    \begin{tikzpicture}[baseline={([yshift=-.5ex]current bounding box.center)}, scale=0.7]
 \pgftext{\includegraphics[scale=1]{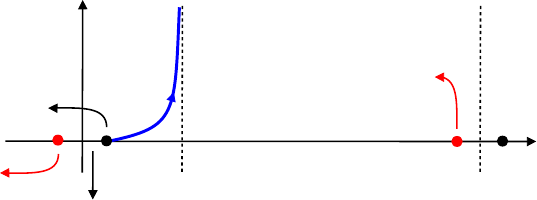}} at (0,0);
    \draw (1.1,-0.05) node {\color{red}horizon};
    \draw (1.1,0.65) node {\color{red}cosmological};
    \draw (-4.8,-0.05) node {horizon};
    \draw (-5.6,-1.25) node {\color{red}interior};
    \draw (-5.6,-1.95) node {\color{red}horizon};
    \draw (4.6,-1.8) node {$r$};
    \draw (3.7,-1.8) node {$2\pi$};
    \draw (-1.3,-1.8) node {$\pi/2$};
    \draw (-3.05,-2.2) node {minus};
    \draw (-3.05,-2.9) node {two-sphere};
    \draw (-3.05,-3.6) node {(section \ref{sect:5.2})};
    \draw (-1.3,2.2) node {\color{blue}holographic screen};
  \end{tikzpicture}\label{2.39 pic}
\end{equation}
The renormalized geodesic length between boundary points on the AdS$_2$ disk is well-known, and gives the following semi-classical approximation for the boundary-to-boundary non-minimally coupled matter correlator \eqref{particle_action} in sine dilaton gravity
\begin{equation}
    e^{-\Delta L}=\frac{\sin(\theta)^{2\Delta}}{\sin(\sin(\theta)\tau/2)^{2\Delta}}\,.\label{2.40 corsine}
\end{equation}
This indeed matches $e^{-2\Delta \varphi}$ in the q-Schwarzian \eqref{ads} provided one used the dictionary
\begin{equation}
    L=2\varphi\,.
\end{equation}
In section \ref{sect:gravdual}, we will show that this dictionary remains true in the full theory, by performing an exact quantization of sine dilaton gravity in the metric formulation. 

\section{Fakeness and length positivity in the q-Schwarzian}\label{sect:fakebdy}
In section \ref{3.1}, following \cite{Goel:2023svz} and \cite{Blommaert:2023wad}, we consider the semi-classical thermodynamics and correlators of DSSYK. As announced in the introduction section \ref{sect:intro}, there are several surprises in this semiclassical regime, including (from a putative bulk point of view) the non-monotonicity of the entropy profile \eqref{1.0 entropy}
\begin{equation}
     S=\frac{\pi \theta-\theta^2}{\abs{\log q}}\,,\label{3.1 ent}
\end{equation}
the distinction between temperature and fake temperature (the decay rate of correlators), and the lack of a UV divergence in correlators. The sine dilaton gravity observables studied in section \ref{2} have none of these features, and reproduce ``fake'' observables. Resolving this mismatch with the bulk theory is an important problem. The question is a version of what UV-completeness looks like in an effective bulk model of quantum gravity. What causes the entropy to decrease? Black holes do not usually have such behavior, their density of states grows with energy. The boundary theory has given us a prediction that is tough to reproduce from the bulk. This is the best possible scenario and is an indication of new rules for the bulk effective gravitational theory. In the remainder of this work, we make progress towards understanding those new ingredients. In particular in section \ref{sect:3.3 new} we point out that at the level of the q-Schwarzian, all mismatches with DSSYK are resolved upon imposing the positivity constraint \eqref{1.7 constraint}
\begin{equation}
    \varphi\geq 0\,,
\end{equation}
and using initial and final conditions $\varphi=0$ for the path integral \eqref{1.20 bis}. This is inspired by the relation \eqref{1.19 dic} with chords in DSSYK $\varphi=\abs{\log q} n$. In particular, we will modify the entropy calculation of the q-Schwarzian of section \ref{2.2}, and show that upon imposing these constraints the q-Schwarzian entropy matches with the DSSYK answer \eqref{3.1 ent}; the correlators now also match. 

\subsection{DSSYK semiclassics}\label{3.1}
We now present the semi-classical analysis of the exact results for the partition function and two-point point in DSSYK, following similar analyses in \cite{Goel:2023svz} and \cite{Blommaert:2023wad}. We start with the known  DSSYK partition function \cite{Berkooz:2018jqr,Berkooz:2018qkz}
\begin{equation}
\label{labelless}
    Z_\text{DSSYK}(\beta)=\int_0^\pi \d\theta\, (e^{\pm 2\i\theta};q^2)_\infty\,\exp\bigg(\beta \frac{\cos(\theta)}{2\abs{\log q}}\bigg)\,,
\end{equation}
and are interested in the regime $\abs{\log q}\ll 1$. Using
\begin{equation}
    \log (x;q^2)_\infty \approx -\frac{\text{Li}_2(x)}{2\abs{\log q}}+\text{subleading}\,,\quad \text{Li}_2(x)+\text{Li}_2(1/x)=-\frac{\pi^2}{6}-\frac{1}{2}\log^2(-x)\,,\label{Li2}
\end{equation}
one obtains the semi-classical entropy
\begin{equation}
    \log\, (e^{\pm 2\i\theta};q^2)_\infty\approx \frac{\pi\theta-\theta^2}{\abs{\log q}}+\text{constant}\,.
\end{equation}
The correct one-loop factor can be extracted from q-Stirling. One arrives (up to overall constants) at the semi-classical DSSYK partition function \cite{Goel:2023svz}
\begin{equation}\label{semicl_partition_function}
    Z_\text{DSSYK}(\beta)\overset{\text{class}}{=} \int_0^\pi \d\cos(\theta)\,\exp\bigg( \frac{\pi\theta-\theta^2}{\abs{\log q}} +\beta \frac{\cos(\theta)}{2\abs{\log q}}\bigg)\,.
\end{equation}
From this one reads off the semi-classical entropy \eqref{3.1 ent} and the (inverse) temperature of DSSYK
\begin{equation}\label{true_temperature}
    \beta=\frac{2\pi-4\theta}{\sin(\theta)}\,\,.
\end{equation}
This mismatches with the naive (no positivity constraints) sine dilaton gravity \eqref{2.10 td} and q-Schwarzian \eqref{2.28 zqschnaive} answers, because of the $\theta^2$-term in the entropy of semi-classical DSSYK. 

The two-point function of DSSYK which computes the expectation value of $e^{-2\abs{\log q}n(\tau)}$ is \cite{Berkooz:2018qkz}
\begin{equation}
    \frac{1}{Z(\beta)}\int_0^\pi \d\theta_1\,\rho(\theta_1)\int_0^\pi \d\theta_2\,\rho(\theta_2)\exp\bigg((\beta-\tau)\frac{\cos(\theta_1)}{2\abs{\log q}}+\tau \frac{\cos(\theta_2)}{2\abs{\log q}}\bigg)\frac{(q^{4\Delta};q^2)_\infty}{(q^{2\Delta}e^{\pm\i\theta_1\pm\i\theta_2};q^2)_\infty}\,.
\end{equation}
Following \cite{Goel:2023svz} and \cite{Blommaert:2023wad}, we scale variables as
\begin{equation}
    \theta_1=\theta+\a \log q\,,\quad \theta_2=\theta-\a\log q\,.
\end{equation}
In the regime $\abs{\log q}\ll 1$ in which we are interested, with $\theta,\a,\Delta$ finite, one has
\begin{equation}
    \frac{(q^{4\Delta};q^2)_\infty}{(q^{2(\Delta\pm\i\a)};q^2)_\infty} \approx \frac{\Gamma(\Delta\pm \i\alpha)}{\Gamma(2\Delta)}\,.
\end{equation}
Furthermore, using \eqref{Li2} and Taylor expanding
\begin{equation}
    \text{Li}_2\left(e^{2\i\theta}+2\Delta\log q \,e^{2\i\theta}\right) \approx \text{Li}_2\left(e^{2\i\theta}\right)-2\Delta \log q \,\log(1-e^{2\i\theta})\,,
\end{equation}
one finds that in our regime of interest
\begin{equation}
    \frac{1}{(q^{2\Delta}e^{\pm 2 \i\theta};q^2)_\infty} \approx \sin(\theta)^{2\Delta-1}\exp\bigg( -\frac{\pi\theta-\theta^2}{\abs{\log q}} \bigg)\,.
\end{equation}
Redefining $\omega=\alpha \sin(\theta)$ the two-point function thus reduces to
\begin{align}\label{saddle_point_approx}
    &\frac{1}{Z(\beta)}\int_0^\pi \d\cos(\theta)\,\exp\bigg( \frac{\pi\theta-\theta^2}{\abs{\log q}} +\beta \frac{\cos(\theta)}{2\abs{\log q}}\bigg)\,\sin(\theta)^{2\Delta}\int_{-\infty}^{+\infty}\d\omega\,e^{\omega(\beta/2-\tau)}\frac{\Gamma(\Delta\pm \i \omega/\sin(\theta))}{\Gamma(2\Delta)}\nonumber\\&\qquad\overset{\text{class}}{=} \sin(\theta)^{2\Delta}\int_{-\infty}^{+\infty}\d\omega\,e^{\omega(\beta/2-\tau)}\frac{\Gamma(\Delta\pm \i \omega/\sin(\theta))}{\Gamma(2\Delta)}\bigg\rvert_{\beta=\frac{2\pi-4\theta}{\sin(\theta)}}\,.
\end{align}
Here we did the $\theta$-integral by saddle-point methods. Finally doing the $\omega$-integral, one obtains the (known \cite{Maldacena:2016hyu}) semi-classical correlation function of DSSYK:
\begin{equation}\label{semicl_twopoint}
e^{-2\Delta \abs{\log q} n}= \frac{\sin(\theta)^{2\Delta}}{\sin(\sin(\theta)\tau/2+\theta)^{2\Delta}}\,\,.
\end{equation}
This has certain surprising features, and importantly mismatches with the naive q-Schwarzian \eqref{ads} and naive sine dilaton gravity \eqref{2.40 corsine} correlator from the previous section.

\subsection{Temperature versus fake temperature}\label{3.2}
The first surprising feature, especially from a putative dual gravitational point of view  is the distinction between temperature $\beta^{-1}$ and ``fake temperature'' $\beta_\text{BH}^{-1}$ \cite{Lin:2023trc}
\begin{equation}
\beta_\text{BH}=\frac{2\pi}{\sin(\theta)}\neq \beta\,. 
\end{equation}
Physical temperature $\beta^{-1}$ is (inverse) periodicity in Euclidean time. Fake (inverse) temperature $\beta_\text{BH}^{-1}$ is the decay time of correlation functions. Usually these notions are identical. However, if we consider the real-time (two-sided) DSSYK correlator (by inserting $\tau=\beta/2+\i T$), one finds that DSSYK correlators decay at late real times $T$ with \emph{fake} temperature
\begin{equation}
    e^{-2\Delta \abs{\log q} n}= \frac{\sin(\theta)^{2\Delta}}{\cosh(\sin(\theta) T/2)^{2\Delta}}\quad\sim \exp\bigg(-\Delta \frac{2\pi}{\beta_\text{BH}}T\bigg)\,. 
\end{equation}
This suggest that the correlators are probing a horizon with Hawking temperature $\beta_\text{BH}^{-1}$. Upon Fourier transforming the LHS using
\begin{equation}
    e^{-2\Delta \abs{\log q} n}=\sin(\theta)^{2\Delta}\int_{-\infty}^{+\infty}\d\omega\,e^{\i\omega T}\,\frac{\Gamma(\Delta\pm \i \omega/\sin(\theta))}{\Gamma(2\Delta)}\,,
\end{equation}
one reaches the same verdict by reading off the quasi-normal modes from the poles of $\Gamma(\Delta\pm \i \omega/\sin(\theta))$. To emphasize how ``odd'' this behavior is, we could consider $\theta=\pi/2$ such that temperature is \emph{infinite} $\beta=0$, but $\beta_\text{BH}=2\pi$. At infinite temperature, ordinary thermal systems would instantly equilibrate, roughly because the average energy per particle diverges. But in DSSYK the average energy remains finite, even at infinite temperature, resulting in finite characteristic times for all correlation functions \cite{Lin:2022nss}. This behavior is universal for systems with a bounded (or truncated) energy spectrum, which is a version of UV-completeness. What is the bulk dual of this universal phenomena? We make first steps towards answering that question in section \ref{sect:gravdual}.

A second surprising feature, from a bulk point of view, is the lack of a UV-divergence in the DSSYK correlator \eqref{semicl_twopoint} for $\tau\to 0$. There are several potential ways out of this. It might hint at genuine non-locality in the bulk, which may arise for instance from a non-commutative bulk spacetime \cite{Almheiri:2024ayc,Berkooz:2022mfk}, or from stringy physics \cite{Maldacena:2016hyu}. It might also hint at finite-cutoff holography \cite{McGough:2016lol,Gross:2019ach,Iliesiu:2020zld,Griguolo:2021wgy}, without the need for holographic renormalization. Our sine dilaton gravity analysis points however to a different resolution, as we investigate in section \ref{sect:defect}. For now, we would like to point out that this lack of UV-divergence can entirely be attributed to the same $\theta^2$-term in the entropy \eqref{3.1 ent} that is responsible for the difference between temperature and fake temperature. Indeed, if we would naively just modify the entropy term in \eqref{saddle_point_approx} by dropping the $\theta^2$-term, the $\theta$ shift in the denominator of \eqref{semicl_twopoint} would disappear
\begin{align}
    &\frac{1}{Z(\beta)}\int_0^\pi \d\cos(\theta)\,\exp\bigg( \frac{\pi\theta}{\abs{\log q}} +\beta \frac{\cos(\theta)}{2\abs{\log q}}\bigg)\,\sin(\theta)^{2\Delta}\int_{-\infty}^{+\infty}\d\omega\,e^{\omega(\beta/2-\tau)}\frac{\Gamma(\Delta\pm \i \omega/\sin(\theta))}{\Gamma(2\Delta)}\nonumber\\&\qquad \overset{\text{class}}{=} \sin(\theta)^{2\Delta}\int_{-\infty}^{+\infty}\d\omega\,e^{\omega(\beta/2-\tau)}\frac{\Gamma(\Delta\pm \i \omega/\sin(\theta))}{\Gamma(2\Delta)}\bigg\rvert_{\beta=\frac{2\pi}{\sin(\theta)}}=\frac{\sin(\theta)^{2\Delta}}{\sin(\sin(\theta)\tau/2)^{2\Delta}}\,.
\end{align}
This suggests conversely that if one would manage to ``cure'' the entropy, by explaining the shift by $\theta^2$ in \eqref{3.1 ent} from the point of view of the q-Schwarzian and sine dilaton gravity, all interesting features of the correlators (the lack of a UV-divergence, and the difference between temperature and fake temperature) should be explained automatically. Save the thermodynamics, save the world. 

\subsection{Length positivity explains the difference of temperature and fake temperature}\label{sect:3.3 new}
Let us now show that in the q-Schwarzian model all mismatches with DSSYK are solved in one go, by imposing the positivity constraint $\varphi\geq 0$ on the classical solution, and using initial and final conditions $\varphi=0$. Recall that the classical solutions of the q-Schwarzian theory take the form \eqref{ads}
\begin{equation}
  e^{-2 \Delta\varphi}=\frac{\sin(\theta)^{2\Delta}}{\sin(\sin(\theta)\tau/2+c)^{2\Delta}}\,.
\end{equation}
Here we included an integration constant $c$, which represents a simple time shift. Part of the classical solution violates $\varphi\geq 0$. We can find a solution that \emph{does} satisfy the constraint $\varphi\geq 0$ and the boundary conditions $\varphi(0)=\varphi(\beta)=0$ by choosing $c=\theta$, and copying the existing solution for the interval where $\varphi$ is non-negative, as follows:
\begin{equation}
    e^{-2 \Delta\varphi}=\frac{\sin(\theta)^{2\Delta}}{\sin(\sin(\theta)\tau/2+\theta)^{2\Delta}}\,,\quad 0<\tau<\frac{2 \pi-4 \theta}{\sin(\theta)}=\beta \,.\quad\begin{tikzpicture}[baseline={([yshift=-.5ex]current bounding box.center)}, scale=0.7]
 \pgftext{\includegraphics[scale=1]{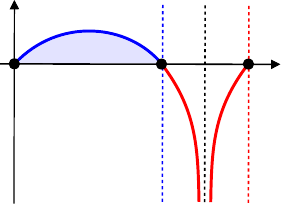}} at (0,0);
    \draw (2.3,0.2) node {$\tau$};
    \draw (-2.6,1.6) node {$\varphi$};
    \draw (-0.9,0.2) node {\color{blue}$\varphi\geq 0$};
     \draw (-0.9,-0.5) node {\color{blue}physical};
    \draw (0.3,2) node {\color{blue}$\beta$};
    \draw (1.9,2) node {\color{red}$\beta_\text{BH}$};
  \end{tikzpicture}\label{3.20 restrsol}
\end{equation}
This precisely matches with the classical DSSYK two point function \eqref{semicl_twopoint}. Indeed, recall $\varphi=\abs{\log q}n$, where $n$ is the number of chords in DSSYK. Crucially, this number of chords must always be positive. The shift by $\theta$ in the denominator ensures furthermore that $n(0)=n(\beta)=0$. 

Here we can also discern the disparity between fake and true temperature, associated with different periodicities of the q-Schwarzian classical solutions depending on whether or not we impose $\varphi\geq 0$.

The remaining task is to check that the q-Schwarzian thermodynamics \emph{with} these constraints indeed also reproduces DSSYK classical thermodynamics \eqref{semicl_partition_function}. The energy of the solution \eqref{2.22 en} is trivially still unchanged, after all the solution \eqref{3.20 restrsol} is a restricted version of the solution considered in section \ref{2.2}. The non-trivial part is the entropy $S_\text{qsch}$ \eqref{entropy}. In the current case, the $\tau$ integral should only run over the range where the solution \eqref{3.20 restrsol} is non-negative:
\begin{equation}
e^{S_\text{qsch}}=\exp \left(\frac{1}{\abs{\log q}} \int_{0}^{\tau_{\mathrm{final}}} \mathrm{d}\tau \left( k \frac{\mathrm{d}}{\mathrm{d}\tau}\phi\right)\right)\,,\quad \tau_\text{final}=\frac{2 \pi-4 \theta}{\sin(\theta)}=\beta\,.
\end{equation}
Introducing again the complex variable $z=e^{\i \sin(\theta)\tau}$, we obtain the analogue of equation \eqref{f(z)}\footnote{The integrand is obtained from that in \eqref{f(z)} by applying the transformation $z\rightarrow e^{2\i \theta} z$. This transformation effectively rotates the branch cut, which now extends from $z=e^{-4\i \theta}$ to $e^{-2\i \theta}$. This transformation translates the shift with $\theta$ in the denominator of the $\varphi$ solution \eqref{3.20 restrsol} and similarly for $k$.}
\begin{equation}\label{g(z)}
S_{\mathrm{qsch}}=\frac{1}{2\abs{\log q}}\int_{\gamma_1} \d z\, \bigg(\frac{1}{z}+\frac{2 e^{2 \i \theta}}{1-e^{2 \i \theta} z}\bigg) \log \bigg(\frac{e^{-\i \theta}-e^{3 \i \theta}z}{1-e^{2 \i \theta}z}\bigg)=\frac{1}{\abs{\log q}}\int_{\gamma_1} \mathrm{d}z\,g(z)
\end{equation} 
However, unlike in \eqref{f(z)}, the contour $\g_1$ in the complex $z$ plane does not close. This change in contour implements the constraint $\varphi\geq 0$ at the semi-classical level
\begin{equation}
    \begin{tikzpicture}[scale=2.3, thick]
	\draw[fill] (0, 0) circle (.7pt) node[shift=(225:0.4)]{0};
	\draw [decorate,decoration={zigzag,segment length=4,amplitude=2,post=lineto,post length=0}] (0.435682,-0.9001) arc (295.82:327.91:1);
	\node at (0.8,0.8) {{\textcolor{red}{$\gamma_1$}}};
	\node at (0.55,-0.4) {{\textcolor{blue}{$\gamma_2$}}};
	\draw [->] (-1.2,0) -- (1.3,0);
	\draw [->] (0,-1.2) -- (0,1.3);
        \draw (1.1,1.1) node {$z$};
	\draw[red] (1,0) arc (0:295.82:1);
	\draw[blue] (0.435682,-0.9001) -- (1,0);
	\draw[fill] (0.847255,-0.531186) circle (.7pt) node[shift=(330:0.4)]{{$e^{-2i\theta}$}};
	\draw[fill] (0.435682,-0.9001) circle (.7pt) node[shift=(330:0.4)]{{$e^{-4i\theta}$}};
	\end{tikzpicture}
\end{equation}
We can deform the contour $\gamma_1$ into the segment $\gamma_2$ and the residue 
from $z=0$; the latter is responsible for the fake thermodynamics discussed in section \ref{2.2}. Therefore one has:
\begin{equation}
    S_{\mathrm{qsch}}=\frac{\pi\theta}{\abs{\log q}}-\frac{1}{\abs{\log q}}\,\int_{\gamma_2}\d z\, g(z)\,.
\end{equation}
The contribution from the open contour $\gamma_2$ can be explicitly evaluated as:
\begin{equation}\label{I4}
\begin{split}
\int_{\gamma_2}\d z\, g(z) &= \pi\theta-\frac{\pi^2}{12} - \frac{1}{2} \mathrm{Li}_{2}\left(e^{-2i\theta}\right) - \frac{1}{2} \mathrm{Li}_{2}\left(e^{2i\theta}\right)  - \frac12 \log(-e^{2 \i \theta})^2=\pi\theta-\frac{1}{4}\log(-e^{2 \i \theta})^2\,.
\end{split}
\end{equation}
Details of this are presented in appendix \ref{A1}. As a nice feature, we immediately observe the emergence of the same dilog functions that appeared in the semi-classical limit of the exact DSSYK spectral density (discussed in section \ref{3.1}).
In the final equality we made use of the $\mathrm{Li}_{2}$ identity \eqref{Li2}. This results in
\begin{equation}
\boxed{S_\text{qsch}=\frac{\pi\theta-\theta^2-\pi^2/4}{\abs{\log q}}}\,.
\end{equation}
This indeed matches with the semi-classical entropy of DSSYK \eqref{semicl_partition_function}. So, imposing the constraint $\varphi \geq 0$ on the q-Schwarzian indeed yields the missing $\theta^2$ contribution to the entropy, resolving the inconsistency with DSSYK thermodynamics.

\section{The gravitational dual of DSSYK}\label{sect:gravdual}
Thus far we have discussed the equivalence of sine dilaton gravity and the q-Schwarzian \emph{without} constraints in section \ref{2}. In section \ref{sect:fakebdy}, we showed at the semi-classical level (on-shell action) that imposing the constraint $\varphi\geq 0$ in the q-Schwarzian boundary theory produces a match with DSSYK. In this section we present the logical consequence of these two facts, namely that sine dilaton gravity with the positive length constraint $\mathbf{L}\geq 0$ is dual to DSSYK. In section \ref{sect:harlowjafferissine} we present the canonical quantization of sine dilaton gravity, following step by step the logic used to canonically quantize JT gravity in \cite{Harlow:2018tqv}. We will show that this precisely reproduces the auxiliary quantum mechanical description \eqref{1.10 hamil} of DSSYK. In section \ref{sect:defect} we present a semi-classical sine dilaton gravity interpretation of the constraint $L\geq 0$ involving a defect and the associated Lorentzian spacetime \cite{Dong:2022ilf}, which crucially is a \emph{smooth} black hole with Hawking temperature $\beta_{\text{BH}}^{-1}$ equal to the fake temperature of DSSYK \eqref{1.7 fakebeta}.

\subsection{Canonical quantization of sine dilaton gravity}\label{sect:harlowjafferissine}
What do we need to canonically quantize a theory? We need the phase space (including its symplectic form) and a Hamiltonian on this phase space, and we need to pray we can solve the resulting Schr\"odinger equation. For 2d dilaton gravity, the first step is simple, as explained nicely by D. Harlow and D. Jafferis \cite{Harlow:2018tqv}. Phase space is the space of classical solutions (this equals the space of initial conditions). For 2d dilaton gravity, this space is two-dimensional. One coordinate is the area of the black hole horizon, proportional to $\Phi_h$. The second coordinate is the amount $T$ of two-sided Lorentzian time evolution of the TFD state. The dilaton gravity model in question is specified entirely by the relation between the ADM energy and $\Phi_h$. 

To be explicit, we will again consider sine dilaton gravity \eqref{1.1 sinedil} (with rescaled dilaton by $1/2\abs{\log q}$)
\begin{equation}
    \int \dpi g\dpi\Phi\,\exp\bigg( \frac{1}{2\abs{\log q}}\bigg\{\frac{1}{2}\int \d x \sqrt{g}\,\big(\Phi R+2\sin(\Phi)\big)+\text{boundary}\bigg\}\bigg) \, ,
\end{equation}
with classical solution:
\begin{equation}
    \Phi=r\,,\quad \d s^2=-F(r)\d t^2+\frac{1}{F(r)}\d r^2\,,\quad F(r)=2\cos(r)-2\cos(\Phi_h)\,.\label{4.2 met}
\end{equation}
Since $\Phi_h$ explicitly appears as a parameter of these solutions, it is obviously a parameter on phase space. For the second coordinate we should consider the Kruskal extension of the above solution. Different two-sided slices of this solution correspond with different initial conditions for sine dilaton gravity, as explained in Figure 3 in \cite{Harlow:2018tqv}, which we reproduce here
\begin{equation}
    \begin{tikzpicture}[baseline={([yshift=-.5ex]current bounding box.center)}, scale=0.7]
 \pgftext{\includegraphics[scale=1]{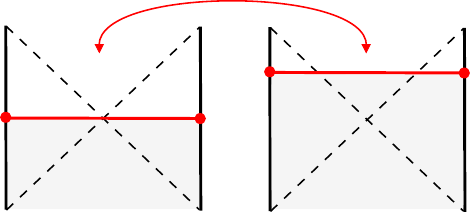}} at (0,0);
    \draw (4.75,0.55) node {$T/2$};
    \draw (-4.5,-0.2) node {$0$};
    \draw (0,4) node {\color{red}geometry \eqref{4.2 met} with identical $\Phi_h$};\draw (0,3.2) node {\color{red}different global slices prepared};
    \draw (0,2.4) node {\color{red}$=$ different initial conditions};
  \end{tikzpicture}
\end{equation}
Using the usual Rindler boost isometry, inequivalent bulk slices (initial conditions) are specified (besides by $\Phi_h$) by the amount of two-sided time evolution $T$
\begin{equation}
    T=T_L+T_R\,.
\end{equation}
Here the bulk slice ends at $t=T_R$ on the right-boundary and at $t=T_L$ on the left. As $\Phi_h$ is independent of time, this leads to the following classical Hamiltonian system (conform equation (2.25) in \cite{Harlow:2018tqv})
\begin{align}
    \dot{T}&=1 \,,\nonumber\\
    \dot{\Phi}_h&=0\, ,\nonumber\\
    H_\text{grav}&=-\frac{\cos(\Phi_h)}{2\abs{\log q}}\,.\label{4.5 ADMham}
\end{align}
The precise ADM Hamiltonian \eqref{4.5 ADMham} for sine dilaton gravity follows unambiguously from the gravitational on-shell action calculated in section \ref{2.1} as the term that multiplies $-\beta$, see in particular equation \eqref{2.bb energy}. So far, the only difference between JT gravity and sine dilaton gravity is the different relation between energy $H$ and entropy $\Phi_h$. From \eqref{4.5 ADMham} one deduces the symplectic form on phase space
\begin{equation}
    \omega=\frac{\sin(\Phi_h)}{2\abs{\log q}}\,\d T\wedge \d \Phi_h=\d T\wedge \d H_\text{grav}\,.
\end{equation}
To be consistent with previous notation, from hereon we will again work with the coordinate $\theta=\Phi_h$.

Sticking to the script of D. Harlow and D. Jafferis, we next introduce more convenient coordinates on this phase space. One natural choice would be to use the length $\ell$ of the geodesic connecting the end points of the spatial slice under consideration (and its canonical conjugate)
\begin{equation}
    \ell=\int \d s\,.
\end{equation}
However, in this case it turns out to be more convenient to work instead with the Weyl-rescaled length variable $L$ \eqref{1.2 matterac}, which can be thought of as the action for matter which is coupled non-minimally to the dilaton. In our current conventions (remember that we rescaled $\Phi$ by $2\abs{\log q}$) this reads
\begin{equation}
    L=\int\d s\,e^{-\i\Phi/2}\,.
\end{equation}
The extremum for this action is a geodesic in the effective metric \eqref{eff}
\begin{equation}
    \d s^2 e^{\i\Phi}=- \sin(\theta)^2\sinh(\rho)^2 \d t^2+\d \rho^2\,.\label{4.9 dseff}
\end{equation}
This is just the Rindler wedge of AdS$_2$ with inverse temperature $\beta_\text{AdS}=2\pi/\sin(\theta)$, so $L$ computes the length of the ERB in a two-sided AdS$_2$ black hole at inverse temperature $\beta_\text{AdS}=2\pi/\sin(\theta)$ with known answer (after appropriate holographic renormalization)
\begin{equation}
    e^{-L}=\frac{\sin(\theta)^2}{\cosh(\sin(\theta)T/2)^2}\,.\label{4.10 L}
\end{equation}
The task is to find the canonical conjugate $p$ to this variable $L$ and find an explicit expression for the Hamiltonian $H_\text{grav}(L,p)$, in terms of these new phase space variables. But how does this variable reflect that we are studying sine dilaton gravity, and not JT gravity? This shines through in the way that $L$ depends on ADM energy $E$ (via its $\theta$ dependence) as compared to the JT gravity expression
\begin{equation}
    e^{-L}=\frac{1-4 \abs{\log q}^2 E^2}{\cosh\Big(\left(1-4 \abs{\log q}^2 E^2\right)^{1/2}T/2\Big)^2}\quad \text{versus}\quad e^{-L_\text{JT}}=\frac{E}{\text{cosh}(E^{1/2}T/2)^2}\,.
\end{equation}
Upon inverting this relation (and the expression for $p(E,T)$) this results in a very different Hamiltonian $H(L,p)=E$ and consequently also different quantum mechanics. Another question is why we consider the variable $L$ over $\ell$? Ultimately, the answer is \emph{because it works}, however there are two a priori reasons.
\begin{enumerate}
    \item As we have discussed in section \ref{sect:operatordic} and section \ref{3.1}, q-Schwarzian and DSSYK correlators match at the classical level with expectation values of $e^{-\Delta L}$ (modulo the temperature vs fake temperature puzzle, which we will address) so that classically $L=2\abs{\log q} n$. This makes it natural to attempt canonical quantization of the bulk using this same $L$ variable, since DSSYK canonical quantization \eqref{1.10 hamil} is naturally phrased using the $n$ variable. 
    \item Canonical quantization of dilaton gravity using $\ell$ generically does not allow to solve for an explicit Hamiltonian function $H(\ell,\pi)$. See equation (4.4) in \cite{Kruthoff:2024gxc}. A potentially more serious problem is that for general potentials there is not one unique geodesic connecting any two boundary points, resulting in an ambiguous Hilbert space even at disk level \cite{Kruthoff:2024gxc}. Such ambiguities are resolved by using a Weyl-transformed length $L$ that effectively computes (unique) geodesics in AdS$_2$.\footnote{We are currently investigating canonical quantization using a Weyl-rescaled $L$ variable for generic dilaton gravity models.}
\end{enumerate}

One can then check that if one defines
\begin{equation}
    e^{-\i P}=-\i \sin(\theta)\tanh(\sin(\theta)T/2)+\cos(\theta)\,,\label{4.12 p}
\end{equation}
that the symplectic form becomes
\begin{equation}
    \omega=\frac{1}{2\abs{\log q}}\d L\wedge \d P=\d T\wedge \d E\,.\label{4.13 omega}
\end{equation}
By inverting the relations for $L$ and $P$ one then finds the Hamiltonian function $H_\text{grav}(L,P)$:
\begin{equation}
    H_\text{grav}=-\frac{\cos(P)}{2\abs{\log q}}+\frac{1}{4\abs{\log q}}e^{iP}e^{-L}\,.\label{4.14 ham}
\end{equation}
Indeed, if one inserts the expressions for $P$ and $L$ in \eqref{4.14 ham}, this indeed reduces to $E=-\cos(\theta)/2\abs{\log q}$. We comment that aside from a direct evaluation of \eqref{4.13 omega}, one way to check the symplectic form $\omega$ is to realize that $P$ and $L$ are solutions to the Hamilton equations with Hamiltonian \eqref{4.14 ham} and symplectic form \eqref{4.13 omega}. On-shell, the symplectic form always reduces to $\d T\wedge \d E$. Indeed\footnote{This is consistent with the fact that entropy $S$ is the integral of $\omega$ enclosed by the orbit with energy $E$, which we know from the on-shell action of the path integral formulation of quantum mechanics.}
\begin{align}
    \omega =\sum_{A} \d x_A(T,E)\wedge \d p_A(T,E)&=\d T\wedge \d E \sum_A \bigg( \frac{\d x_A}{\d T}\frac{\d p_A}{\d E}-\frac{\d x_A}{\d E}\frac{\d p_A}{\d T}\bigg)\\ &=\d T\wedge \d E \sum_A \bigg( \frac{\d H}{\d p_A}\frac{\d p_A}{\d E}+\frac{\d H}{\d x_A}\frac{\d x_A}{\d E}\bigg)=\d T\wedge \d E\,\frac{\d H}{\d E}=\d T\wedge \d E\,.\nonumber
\end{align}

There is in general more than one way to quantize a classical system. For this particular case, we have explored distinct approaches in our follow-up work \cite{Blommaert:2024whf}. The upshot is that one can quantize the system \eqref{4.14 ham} in a ``Liouville'' approach, leading to a dynamical system closely related to the recent complex Liouville string approaches of \cite{Collier:2024kmo}. Another way, which is the one we present here, agrees exactly with DSSYK, which is our main interest. Upon quantization, the Hamiltonian \eqref{4.14 ham} becomes
\begin{equation}
    \boxed{\mathbf{H}_\text{grav}=-\frac{\cos(\mathbf{P})}{2\abs{\log q}}+\frac{1}{4\abs{\log q}}e^{i\mathbf{P}}e^{-\mathbf{L}}}\label{4.16 ham}
\end{equation}
Here, according to the symplectic form $\omega$ \eqref{4.13 omega}, the operators are quantized as $[\mathbf{L},\mathbf{P}]=2\i\abs{\log q}$. This is precisely the auxiliary quantum mechanical model \eqref{1.10 hamil} of DSSYK, with the identification of $L$ with chord number $n$ extending to an operator identification in the full quantum theory
\begin{equation}
    \boxed{\mathbf{L}= 2\abs{\log q}\mathbf{n} }\,.
\end{equation}

With this identification, it is immediate how the dictionary between DSSYK and sine dilaton gravity works at the quantum level. Indeed, directly adopting the results of the auxiliary QM calculation \cite{Berkooz:2018jqr,Berkooz:2018qkz,Blommaert:2023opb} which we reviewed in section \ref{sect:background}, we see that observables in DSSYK map to \emph{transition matrix elements} between states $\ket{\mathbf{L}=0}$, that the quantum theory should be supplemented with the length positivity constraint
\begin{equation}
    \mathbf{L}\geq 0\,.
\end{equation}
We stress that, as shown by the analysis of section \ref{2}, this constraint is not a standard property of sine dilaton gravity. It is a new ingredient in the bulk theory deduced from the boundary side of the duality, required to explain a truncating (UV complete) spectrum such as \eqref{1.0 entropy}.

The DSSYK partition function is computed in sine dilaton gravity as
\begin{equation}
    \text{Tr}\big(e^{-\beta H_\text{SYK}}\big)=\bra{\mathbf{L}=0}e^{-\beta \mathbf{H}_\text{grav}}\ket{\mathbf{L}=0}\,.
\end{equation}
The sine dilaton gravity WdW wavefunctions are found by solving the Schr\"odinger equation associated with the Hamiltonian \eqref{4.16 ham},\footnote{The Hamiltonian \eqref{4.16 ham} is non-Hermitian. Contrary to common intuition this is not a problem, however it does imply that the left-and right eigenvectors of the Hamiltonian are different (as explained in \cite{Blommaert:2023opb}, following \cite{yao2018edge}):
\begin{equation}
    \braket{L\rvert E}=\frac{1}{(q^2;q^2)_{L/2\abs{\log q}}}H_{L/2\abs{\log q}}(-2\abs{\log q} E\,\rvert q^2)\,.
\end{equation}
It can equivalently be mapped into a Hermitian Hamiltonian following \cite{Lin:2022rbf}.
} 
leading to
\begin{equation}
    \braket{E\rvert L}=H_{L/2\abs{\log q}}(-2\abs{\log q} E\,\rvert q^2)\,,
\end{equation}
where $H_n(x\vert q)$ are $q$-Hermite polynomials. In the approach of our follow-up work \cite{Blommaert:2024whf}, the discretization of the length $L$ emerges from gauging the periodicity of the momentum variable $\mathbf{P}$. As a consequence, from the dilaton gravity point of view, the gravitational Hilbert space of sine dilaton gravity is \emph{discretized} (this is completely standard in the DSSYK side of the duality)
\begin{equation}
    \boxed{\mathbb{1}=\sum_{L/2\abs{\log q}=0}^\infty \ket{L}\bra{L}}\,.
\end{equation}
Another way to say this is that \emph{spacetime} is discretized, even though we started with a simple innocent-looking action \eqref{1.1 sinedil} without a priori discretization. Thus, in our theory, discretization is a consequence of quantization, in the same way that in the q-Schwarzian theory non-commutativity (the algebra of the coordinates on the quantum group manifold $(\varphi,\beta,\gamma)$) follows from canonical quantization and is not a priori present in the classical theory \cite{Blommaert:2023opb}. This suggests that perhaps one should think of sine dilaton gravity as identical to quantum gravity on the quantum hyperbolic disk \cite{Almheiri:2024ayc,Berkooz:2022mfk,Berkooz:2023cqc}, with the latter picture arising upon quantization.

More generically, it is remarkable that in this relatively simple model of gravity a discretized bulk Hilbert space arises almost for free, whereas for instance in JT gravity one has to work very hard \cite{Iliesiu:2024cnh}, including wormholes \cite{Cotler:2016fpe,Saad:2018bqo,Saad:2019pqd,Blommaert:2019hjr,Almheiri:2019qdq,Penington:2019kki,Iliesiu:2021ari, Blommaert:2020seb,Blommaert:2022lbh,Saad:2022kfe,Griguolo:2023jyy,Griguolo:2024htx} and modify the definition of the theory to reduce it to one member of the ensemble \cite{Blommaert:2019wfy,Marolf:2020xie,Blommaert:2021fob,Saad:2019lba} to obtain a discretized Hilbert space.

Operators $\mathcal{O}_\Delta$ in DSSYK correspond to insertions of
\begin{equation}
   \mathcal{O}_\Delta \, \leftrightarrow \, q^{2\Delta \mathbf{n}}=e^{-\Delta \mathbf{L}}\,.
\end{equation}
The latter, as emphasized earlier, corresponds in sine dilaton gravity with shooting in non-minimally coupled probe matter \eqref{1.2 matterac} from the holographic boundary. The two-point correlation function of operators $\mathcal{O}_\Delta$ is computed in gravity as
\begin{equation}
    \text{Tr}\big(e^{-\tau H_\text{SYK}}\mathcal{O}_\Delta e^{-(\beta-\tau) H_\text{SYK}}\mathcal{O}_\Delta\big)=\bra{\mathbf{L}=0}e^{-\tau \mathbf{H}_\text{grav}}\,e^{-\Delta \mathbf{L}}\,e^{-(\beta-\tau)\mathbf{H}_\text{grav}} \ket{\mathbf{L}=0}\,.
\end{equation}
Higher point functions with non-crossing lines work similarly.

\subsection{Defect interpretation of the length positivity constraint}\label{sect:defect}
We have established that sine dilaton gravity is exactly equivalent to DSSYK, as long as we impose the constraint $\mathbf{L}\geq 0$ \eqref{1.7 constraint} on the theory and compute transition matrix elements between states $\ket{\mathbf{L}=0}$. As clarified already at the level of the (boundary) q-Schwarzian theory in section \ref{sect:fakebdy}, these two elements are responsible for the distinction between temperature and fake temperature:
\begin{equation}
    \beta=\frac{2\pi-4\theta}{\sin(\theta)}\quad \text{versus}\quad \beta_\text{BH}=\frac{2\pi}{\sin(\theta)}\,,\label{4.28 temps}
\end{equation}
for the non-monotonicity of DSSYK entropy \eqref{1.0 entropy}, and for the apparent non-locality in bulk correlators, namely the absence of short-Euclidean-time divergence in \eqref{semicl_twopoint}.

Even though an exact quantum description is great, it does not address every question we can ask. What have we really learned in terms of universal features of UV-complete (bounded spectrum) simple bulk gravity models? In similar spirit, the fact that the entropy has a maximum may be considered a smoking gun for emergent dS physics \cite{Dong:2018cuv,Chandrasekaran:2022cip}. This is closely tied with a need to introduce observers as part of the quantum description of dS space. At this point, it is not clear how (if at all) this emerges from imposing $\mathbf{L}\geq 0$ (and the initial states) at the quantum level.

Here we will take a first step in this direction. In particular, we present a semi-classical bulk picture within sine dilaton gravity that mimics the constraint $\ket{\mathbf{L}=0}$ and explains in a clear and tangible way why (semi-classical) correlators appear to probe a black hole at \emph{fake} temperature $1/\beta_\text{BH}$. We will first propose an ansatz for such a picture for the partition function, and then show that this same ansatz correctly predicts all semi-classical correlators. We discuss future steps in section \ref{sect:discnew}. 

We propose to modify the gravitational path integral by inserting a ``defect'' operator, similar as in the JT context \cite{Mertens:2019tcm}, with an opening angle $\gamma$:
\begin{equation}
    \mathcal{V}_\gamma=\int\d x \sqrt{g}\,e^{-(2\pi-\gamma)\Phi}\,.
    \label{4.29 defects}
\end{equation}
This creates a source of curvature in classical spacetime at $x=0$,\footnote{Unlike in JT gravity \cite{Mertens:2019tcm}, this delta-source of curvature is only exact at the classical level.}
\begin{equation}
    \sqrt{g}R \,\supset \, 2(2\pi-\g)\delta(x)\,.
\end{equation}
For any fixed $\gamma$ (independent of $\theta$) the on-shell action calculation of section \ref{2.1} gets modified classically to
\begin{equation}
    \begin{tikzpicture}[baseline={([yshift=-.5ex]current bounding box.center)}, scale=0.6]
 \pgftext{\includegraphics[scale=1]{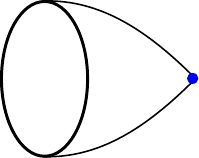}} at (0,0);
    \draw (2.2,0) node {\color{blue}$\mathcal{V}_\gamma$};
    \draw (-2.1,0.5) node {$\beta$};
  \end{tikzpicture}\overset{\text{class}}{=} \int \d E( \theta)\,\exp\bigg(\frac{\gamma\,\theta}{2\abs{\log q}}+\beta \frac{\cos(\theta)}{2\abs{\log q}}\bigg)\,,\label{4.31 defect}
\end{equation}
resulting in modified thermodynamics $\beta=\gamma/\sin(\theta)$. From \eqref{4.28 temps}, it is clear that one can reproduce the correct DSSYK thermodynamics by choosing $\gamma=2\pi-4\theta$ \eqref{4.29 defects}. This is misleading, because $\gamma$ and $\theta$ are conjugate variables in the gravitational path integral \cite{Yang:2018gdb}, so it would be wrong to simply replace $\gamma$ by its $\theta$-dependent value in \eqref{4.31 defect}. Instead, the correct procedure is to work with a \emph{wavefunction} on the opening angle $\gamma$:
\begin{equation}
   \text{Tr}\big(e^{-\beta H_\text{SYK}}\big)=\int \d\gamma\,\psi(\gamma)\quad\begin{tikzpicture}[baseline={([yshift=-.5ex]current bounding box.center)}, scale=0.6]
 \pgftext{\includegraphics[scale=1]{dssyktalkCERN5.pdf}} at (0,0);
    \draw (2.2,0) node {\color{blue}$\mathcal{V}_\gamma$};
    \draw (-2.1,0.5) node {$\beta$};\label{4.32 part}
  \end{tikzpicture}
\end{equation}
In particular, in order to obtain $\gamma=2\pi-4\theta$ on the saddle, one should work with a Gaussian wavefunction
\begin{equation}
    \text{Tr}\big(e^{-\beta H_\text{SYK}}\big)\overset{\text{class}}{=}\int \d\gamma\, \exp\bigg(\frac{(2\pi-\g)^2}{16\abs{\log q}}\bigg)\int \d E(\theta)\,\exp\bigg(\frac{\gamma \theta}{2\abs{\log q}}+\beta \frac{\cos(\theta)}{2\abs{\log q}}\bigg)\,.\label{4.33 def}
\end{equation}
Before proceeding to the Lorentzian interpretation of these spacetimes (and the associated Lorentzian correlators) we pause to make two comments.
\begin{enumerate}
    \item Surprisingly, a mildly modified version of equation \eqref{4.33 def} is exact. Indeed, following equation (64) in \cite{Verlinde:2024znh} one can rewrite the exact DSSYK partition function \eqref{labelless} as
    \begin{align}
        &\text{Tr}\big(e^{-\beta H_\text{SYK}}\big)=\int \d\gamma\, \exp\bigg(\frac{(2\pi-\g)^2}{16\abs{\log q}}\bigg)\sum_{n=-\infty}^{+\infty}\int \d E(\theta)\,\sinh\bigg(\frac{\gamma (\theta+2\pi n)}{2\abs{\log q}}\bigg)\exp\bigg(\beta \frac{\cos(\theta)}{2\abs{\log q}}\bigg)\nonumber\\ &=\int\d E(\theta)\,\rho(E)\,\exp\bigg(\beta \frac{\cos(\theta)}{2\abs{\log q}}\bigg)\,,\qquad \rho(E)=\sum_{n=-\infty}^{+\infty} (-1)^n e^{-\frac{1}{\abs{\log q}}\left(\theta+\pi(n-\frac12)\right)^2}\,.\label{4.34 magic}
    \end{align}
    The sum over $n$, as well as the two branches of the sinh, corresponds according to \cite{Kruthoff:2024gxc} to summing over distinct saddle-points in the gravitational path integral, where each saddle corresponds with spacetime ending on a different horizon. Indeed, in sine dilaton gravity as explained around \eqref{2.bb horizons} there is an infinite series of horizons. There are black hole horizons at (rescaled) $\Phi=\theta+2\pi n$ and cosmological (or interior) horizons at $\Phi=-(\theta+2\pi n)$, which contribute with a (not well understood) minus sign \cite{Kruthoff:2024gxc}. The defect for each saddle sits at the extremum of the dilaton, thus at the horizon in question. We find it baffling that \eqref{4.34 magic} is exact, and we hope to return to this in the future.
    \item In the circular WdW slicing used for instance in \cite{Iliesiu:2020zld}, we can think of this wavefunction in \eqref{4.33 def} as changing the usual $\gamma=2\pi$ no-boundary Hartle-Hawking state in quantum gravity. Apparently in DSSYK there is a fundamental uncertainly in the boost angle represented by a slightly smeared distribution around $\g=2\pi$. The physical meaning is unclear to us.
    
\end{enumerate}

We will now discuss the Lorentzian interpretation of our partition function \eqref{4.32 part}. To this end, we decompose the semi-classical defect path integral in the fixed area basis \cite{Dong:2022ilf,Blommaert:2023vbz}. The defect correlation function can be expanded as \cite{Dong:2022ilf,Blommaert:2023vbz}
\begin{align}
    &\average{ e^{-(2\pi-\gamma)\Phi(x)} }_\beta = \int_{-\infty}^{+\infty} \d A\, e^{A(\g-\a)} \\&\qquad \times\int_{-\i\infty}^{+\i\infty} \d \a\, \int \dpi g\dpi \Phi \exp\bigg(-(2\pi-\alpha)\Phi(x)+\frac{1}{2}\int \d x \sqrt{g}\bigg(\Phi R+\frac{\sin(2\abs{\log q} \Phi)}{\abs{\log q}}\bigg)+\text{boundary}\bigg)\,.\nonumber
\end{align}
The path integral on the second line has a classical saddle for
\begin{equation}
    \alpha=\frac{1}{2}\beta \frac{\sin(2\abs{\log q} \Phi_h)}{\abs{\log q}}\,.\label{4.36 saddle}
\end{equation}
Here we've used the fact that the $\alpha$ variation forces $A=\Phi(x)$ and that classically the defect is located at the extremal surface (where the area is minimal) $\Phi(x)=\Phi_h$ \cite{Blommaert:2023vbz}. The integrals over $A$ and $\a$ reduce, after imposing the saddle \eqref{4.36 saddle}, to only an integral over $A$. Therefore the defect correlator is
\begin{equation}
    \average{ e^{-(2\pi-\gamma)\Phi(x)} }_\beta=\int_{-\infty}^{+\infty} \d A\,e^{-(2\pi-\g) A}\,Z(\beta,A)\,,
\end{equation}
where the fixed area path integral is classically
\begin{equation}
    Z(\beta,A)\overset{\text{class}}{=}\exp\bigg( 2\pi A + \beta \frac{\cos(2\abs{\log q}A)}{2\abs{\log q}}\bigg)\,.
\end{equation}
Introducing a Gaussian wavefunction which suppresses large areas $A\sim 1/\abs{\log q}$
\begin{equation}
    \psi(A)=\int \d \gamma\, \psi(\gamma)\,e^{-(2\pi-\gamma)A}\,,
\end{equation}
one may thus interpret the DSSYK partition function \eqref{4.32 part} as decomposing into fixed area states (or path integrals) as follows
\begin{equation}
   \text{Tr}\big(e^{-\beta H_\text{SYK}}\big)=\int_{-\infty}^{+\infty} \d A\,\psi(A)\quad\begin{tikzpicture}[baseline={([yshift=-.5ex]current bounding box.center)}, scale=0.6]
 \pgftext{\includegraphics[scale=1]{dssyktalkCERN5.pdf}} at (0,0);
    \draw (2.2,0) node {\color{blue}$A$};
    \draw (-2.1,0.5) node {$\beta$};\label{4.39 fixeda}
  \end{tikzpicture}
\end{equation}
This decomposition is useful for our purposes because the Lorentzian interpretation of fixed area states (or fixed area path integrals) has been studied in great detail \cite{Dong:2022ilf}, which leads straight to a Lorentzian interpretation for our defect partition function \eqref{4.32 part}. This picture, obtained in appendix B of \cite{Dong:2022ilf}, is rather beautiful.\footnote{They considered JT gravity, however the generalization to arbitrary dilaton potentials is straightforward.} It turns out that \emph{half} of the defect is part of the Euclidean preparation for the bra, whereas the other half of the singular source is part of the Euclidean preparation for the ket. Crucially, the associated two-sided Lorentzian analytic continuation is a completely smooth spacetime: 
\begin{equation}
    \begin{tikzpicture}[baseline={([yshift=-.5ex]current bounding box.center)}, scale=0.7]
 \pgftext{\includegraphics[scale=1]{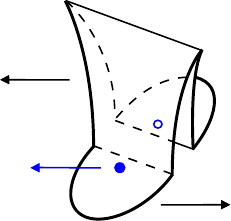}} at (0,0);
    \draw (2.5,0.2) node {$\beta/2$};
     \draw (-2.9,-0.9) node {\color{blue}half defect};
    \draw (4.2,-1.6) node {state preparation};
    \draw (-4.4,0.6) node {Lorentzian smooth};
  \end{tikzpicture}\label{4.41 half}
\end{equation}
To be more precise, the Lorentzian analytic continuation for fixed $A$ (or fixed $\theta$ adhering to our previous notation) is exactly the spacetime one would find by analytically continuing the smooth Euclidean disk with temperature $\beta(\theta)$, fine-tuned to this area $\theta$. In particular, for fixed $\theta$ the Lorentzian spacetime is the analytic continuation of \eqref{metrrr} with
\begin{equation}
    F(r)=-2\cos(r)+2\cos(\theta)\,,\quad 0<\tau<\frac{2\pi}{\sin(\theta)}=\beta_\text{BH}\,.
\end{equation}
So, the Lorentzian space is a smooth two-sided black hole whose Hawking temperature equals the fake temperature $1/\beta_\text{BH}$ \eqref{1.7 fakebeta}! This interpretation of the DSSYK partition function creates the \emph{expectation} that real-time DSSYK correlators should simply map to QFT correlators probing this Lorentzian black hole. Therefore, if we consider the following two-sided matter correlators \eqref{1.2 matterac}
\begin{equation}
    e^{-\Delta L}\,,\quad L=\int\d s\,e^{-\i\abs{\log q}\Phi}\,,
\end{equation}
then the sine dilaton gravity defect picture predicts that at fixed $\theta$ the DSSYK two-point function should be
\begin{equation}
    \begin{tikzpicture}[baseline={([yshift=-.5ex]current bounding box.center)}, scale=0.7]
 \pgftext{\includegraphics[scale=1]{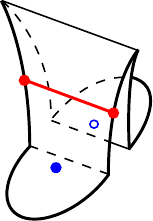}} at (0,0);
    \draw (1.75,-0.35) node {\color{red}$\mathcal{O}_\Delta$};
    \draw (-1.7,0.9) node {\color{red}$\mathcal{O}_\Delta$};
  \end{tikzpicture}\,=\frac{\sin^{2\Delta}(\theta)}{\cosh^{2\Delta}(\sin(\theta)T/2)}\,.\label{4.43 pic}
\end{equation}
Comparing with the semi-classical DSSYK correlator \eqref{semicl_twopoint}, we find that this indeed matches precisely for $\tau=\beta/2+\i T$. For the four-point function of pairwise distinct operators, similarly, the sine dilaton gravity defect picture predicts
\begin{equation}
    \begin{tikzpicture}[baseline={([yshift=-.5ex]current bounding box.center)}, scale=0.7]
 \pgftext{\includegraphics[scale=1]{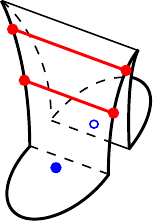}} at (0,0);
    \draw (1.85,0.45) node {\color{red}$\mathcal{O}_{\Delta_2}$};
    \draw (1.85,-0.35) node {\color{red}$\mathcal{O}_{\Delta_1}$};
    \draw (-1.8,1.7) node {\color{red}$\mathcal{O}_{\Delta_2}$};
    \draw (-1.8,0.9) node {\color{red}$\mathcal{O}_{\Delta_1}$};
  \end{tikzpicture}\,=\frac{\sin(\theta)^{2\Delta_1}}{\cosh(\sin(\theta) T_1/2)^{2\Delta_1}}\,\frac{\sin(\theta)^{2\Delta_2}}{\cosh(\sin(\theta) T_2/2)^{2\Delta_2}}\,.\label{4.44 pic}
\end{equation}
We check in appendix \ref{app:DSSYK4pt} that this matches indeed with the semi-classical limit of the DSSYK four-point function. 

One can also make predictions for correlators separated by Euclidean times, by analytically continuing the geodesics in \eqref{4.43 pic} and \eqref{4.44 pic}. For instance for the two-point function this leads to
\begin{equation}
    \begin{tikzpicture}[baseline={([yshift=-.5ex]current bounding box.center)}, scale=0.7]
 \pgftext{\includegraphics[scale=1]{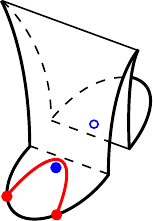}} at (0,0);
    \draw (0.1,-2.25) node {\color{red}$\mathcal{O}_\Delta$};
    \draw (-1.2,-2.1) node {\color{red}$\mathcal{O}_\Delta$};
  \end{tikzpicture}\quad=\frac{\sin^{2\Delta}(\theta)}{\sin^{2\Delta}(\sin(\theta)\tau/2+\theta)}\,.\label{4.46 analyticcont}
\end{equation}
The computation which was carried out here is that of the geodesic length $L$ in a (effective AdS$_2$) defect geometry with deficit angle $\gamma_\text{half}=2\pi-2\theta$. Indeed, the Euclidean preparation region is a patch of the Euclidean disk with circumference $2\pi/\sin(\theta)$ in which one identifies two geodesics emanating from the center of the disk with angle $2\theta$ (\emph{half} the deficit of the partition function \eqref{4.32 part}, as discussed in \eqref{4.41 half}). One way to quickly get the answer is to consider the expression for the boundary correlator in a defect background, which according to for instance equation (4.15) in \cite{Mertens:2019tcm} is
\begin{equation}
    \sum_{w=-\infty}^{+\infty}\frac{\sin^{2\Delta}(\theta)}{\sin^{2\Delta}(\sin(\theta)\tau/2-w(\pi-\theta))}\,.
\end{equation}
From the picture \eqref{4.46 analyticcont} we see that we should be interested in the $w=1$ geodesic, as we are winding once around the half-defect. This reproduces the claimed answer. This exactly reproduces the DSSYK semi-classical two-point function \eqref{semicl_twopoint}, which is a nontrivial check on our semi-classical defect interpretation of the constraint $\mathbf{L}\geq 0$ (and the initial states $\ket{\mathbf{L}=0}$) in the quantization of sine dilaton gravity.

Before closing this work let us make a few more comments regarding this classical gravity picture.
\begin{enumerate}
    \item The geometry in \eqref{4.46 analyticcont} suggests (correctly) that the half-defect replaces the usual no-boundary state preparation at $\tau=0$, as the particle winds around the half-defect. The equation then shows that this state is  $\ket{\mathbf{L}=0}$. Indeed, the minimal length $L$ that the analytically continued geodesic wrapping around the half-defect can attain is $L=0$.
    \item This semi-classical sine dilaton gravity picture is close in spirit to the ``fake disk'' \cite{Lin:2023trc}, especially since the non-minimally coupled matter \eqref{1.2 matterac} can be viewed as probing an effective hyperbolic disk \eqref{eff}. Therefore, in some sense the fake disk is a \emph{real} part of the bulk description in sine dilaton gravity. It would be interesting to understand how the SU$_q(1,1)$ quantum group symmetries \cite{Berkooz:2022mfk,Blommaert:2023opb,Lin:2023trc,Almheiri:2024ayc} are interpreted geometrically in terms of our dilaton gravity (aside from the Poison-sigma model interpretation \cite{Blommaert:2023wad}).
    \item At a semi-classical bulk level, one can also predict a version of the semi-classical OTO four-point function, which would corresponds simply with crossing $\Delta_1$ and $\Delta_2$ lines in the Lorentzian region. This should reproduce the known KummerU-function  \cite{Maldacena:2016upp,Lam:2018pvp}. This leads to the scrambling time $T_\text{S}=\log S/\sin(\theta)$ and a Lyapunov exponent
    \begin{equation}
        \lambda_\text{L}=\sin(\theta)=\frac{2\pi}{\beta_\text{BH}}\,.
    \end{equation}
    This is indeed the Lyapunov exponent for DSSYK in the semi-classical limit \cite{Maldacena:2016hyu,Streicher:2019wek,Choi:2019bmd}. From the boundary point of view, this is \emph{sub-maximal} scrambling \cite{Maldacena:2015waa}, since $\lambda_\text{L}<2\pi/\beta$. But it is actually \emph{maximal} scrambling from a bulk point of view, because as we have seen the dual bulk geometries are Lorentzian black holes with Hawking temperature equal to the fake temperature $1/\beta_\text{BH}$.
\end{enumerate}

\section{Sine dilaton gravity and Liouville de Sitter gravity}\label{sect:5.1}

In this section we show that classically sine dilaton gravity with our choice of boundary conditions and counterterm \eqref{b+b} can be rewritten as a non-critical string consisting of two 2d Liouville actions with complex conjugate central charges. This Liouville system has been advocated by H. Verlinde \cite{HVerlindetalk} to describe the DSSYK model, and to be interpretable as a theory of dS gravity \cite{Verlinde:2024zrh,Verlinde:2024znh}. It would be interesting to compare this in more detail to our work. The interpretation of the positivity constraint \eqref{1.7 constraint} (and whether that is related with inserting observers in dS) in this context in particular would be interesting to understand. 

What follows is morally an analytic continuation of the story for sinh dilaton gravity \cite{Mertens:2020hbs,Fan:2021bwt}. Consider the dilaton gravity action \eqref{b+b} with $g_{\mu\nu}=e^{2 \rho} \delta_{\mu\nu}$. Introducing furthermore
\begin{equation}
    b\phi=\rho+\i\abs{\log q}\Phi\,,\quad \i b\chi=\rho-\i\abs{\log q}\Phi\,,\quad \pi b^2=\i\abs{\log q}\,,
\end{equation}
one obtains (with in this subsection indices referring to the flat fiducial metric $\delta_{\mu\nu}$)
\begin{equation}
\frac{1}{2}\int \d^2 x \sqrt{g}\Phi R+\int \d\tau \sqrt{h} \Phi K=\int \d^2 x\,\partial^\mu \Phi\, \partial_\mu \rho=\int \d^2x\bigg( \frac{1}{4\pi}\partial^\mu \phi\partial_\mu\phi+\frac{1}{4\pi}\partial^\mu \chi\partial_\mu\chi\bigg)\,.
\end{equation}
Furthermore the sine potential becomes
\begin{equation}
\frac{1}{2}\int \d^2 x\,\frac{\sin(2\abs{\log q}\Phi)}{\abs{\log q}}=\int \d^2 x\bigg(-\frac{\i}{4\abs{\log q}}e^{2 b \phi}+\frac{\i}{4\abs{\log q}}e^{\i 2 b \chi}\bigg)\,.
\end{equation}
The counterterm required for holographic renormalization becomes
\begin{equation}
    \int \d\tau \sqrt{h} \bigg(-\i\,\frac{e^{-\i \abs{\log q}\Phi}}{2\abs{\log q}}\bigg)=\int \d\tau\bigg(-\frac{\i}{2\abs{\log q}}e^{\i b\chi}\bigg)\,,
\end{equation}
and the boundary condition \eqref{2.3 bc} is identified with a fixed length boundary condition on the $\phi$ field
\begin{equation}
    e^{b\phi}=\frac{\i}{2\abs{\log q}}\,.
\end{equation}
In total we have two copies of Liouville CFT, each of which has the classical action and central charge
\begin{equation}
    \int \d^2x\bigg( \frac{1}{4\pi}\partial^\mu \psi\partial_\mu\psi+\mu e^{2 b\psi}\bigg)+\int\d \tau \bigg(\mu_{B}e^{b\psi}\bigg)\,,\quad c=1+6\bigg(b+\frac{1}{b}\bigg)^2\,.
\end{equation}
The fields $\phi$ and $\chi$ are distinguished by their (complex) value of $b$ which differs by a factor of $\i$, because of this difference one finds that they have complex conjugate central charges
\begin{equation}
c_{\phi}=13+6 \i \left(\frac{\pi}{\abs{\log q}}-\frac{\abs{\log q}}{\pi}\right)\,,\quad c_{\chi}=13-6 \i \left(\frac{\pi}{\abs{\log q}}-\frac{\abs{\log q}}{\pi}\right)\,,\quad c_\phi+c_\chi=26\,.
\end{equation}
Their bulk cosmological constants are also complex conjugates
\begin{equation}
\label{eq:mus}
    \mu_\phi=-\frac{\i}{4\abs{\log q}}\,,\qquad \mu_\chi=\frac{\i}{4\abs{\log q}}\,.
\end{equation}
As the $\phi$ field is subject to fixed length boundary conditions, it has no boundary cosmological constant in the action. Indeed, upon going to fixed length boundary conditions one inverse Laplace transforms with respect to $\mu_{\phi B}$ hence removing this boundary term \cite{Mertens:2020hbs}. 

We remark that the Liouville metric associated to the field $\chi$ ($\sim$ AdS$_2$) blows up near the holographic boundary, akin to ZZ boundary conditions in ordinary Liouville CFT \cite{Zamolodchikov:2001ah}. 
For the $\chi$ field, we read off the boundary cosmological constant:
\begin{equation}
\label{eq:ZZ}
    \mu_{\chi B}=-\frac{\i}{2\abs{\log q}}\,.
\end{equation}
This again heuristically matches classically with ZZ-brane boundary conditions as follows. It is an old fact in spacelike Liouville CFT that $(m,n)$ ZZ-branes can be viewed as special cases of FZZT branes with a complexified brane parameter \cite{Martinec:2003ka} of the type:
\begin{equation}
\mu_B(m,n) = (-)^m \sqrt{\mu} \frac{\cos (n \pi b^2)}{\sqrt{\sin \pi b^2}}\,.
\end{equation}
Reinstating factors of $\hbar$ (essentially by $b^2 \to \hbar b^2$), and using our specific value for the bulk $\mu$ \eqref{eq:mus}, we obtain for the classical $\hbar \to 0$ limit
\begin{equation}
\mu_B(1,1) \approx - \frac{1}{2\abs{\log q}}\,,
\end{equation}
which is to be compared to \eqref{eq:ZZ}.

We note also that this string theory model is an analytic continuation of the Virasoro minimal string \cite{Collier:2023cyw}, which is a specific model of Liouville gravity.\footnote{Here ``analytic continuation'' is an abuse of language. The parameter $b$ is analytically continued but the amplitudes do not depend on $b$ analytically because the solutions to the bootstrap equations change significantly.} The same authors have modified their techniques to derive the higher genus amplitudes of Liouville dS gravity \cite{Collier:2024kmo}, which would provide one possible non-perturbative definition of sine dilaton gravity. Another possibility perhaps, if one figures out how to implement the discretization constraint (such that the spectrum truncates), is that this could match a finite-cut matrix integral in the large $N$ limit (as proposed in our follow-up work \cite{Blommaert:2025avl}).

Within the coupled Liouville CFT model, the natural operator dictionary would be to map to the open string tachyons \cite{knizhnik1988fractal}
\begin{equation}
    \mathcal{B}_{\beta_M}\sim \int_{\partial\Sigma}\d x\,e^{\i b\Delta\chi }e^{(b-b\Delta )\phi}.
\end{equation}
At least on the classical level this works. The classical solutions for $\chi$ and $\phi$ are just AdS$_2$ Weyl factors, so one would be computing a CFT two-point function on a hyperbolic disk, which reproduces the sine dilaton gravity correlator \eqref{2.40 corsine}.\footnote{We thank Vladimir Narovlansky for discussions on this point.} Of course, this still mismatches with DSSYK, as we did not impose the positivity constraint in this Liouville CFT language. It would be very interesting to understand how to do this, especially at the quantum level. 

\section{Concluding remarks}\label{sect:discnew}
We have presented an exact holographic duality between sine dilaton gravity and DSSYK. Many open questions remain. In particular, the distinction between temperature and fake temperature seems to be a universal feature of systems for which the spectrum truncates (and the entropy has a maximum). This is a version of UV-completeness. Indeed, such systems will always have finite average excitation energies even when the physical temperature is infinite $\beta=0$. This results in finite characteristic times for all physical processes \cite{Lin:2022nss}, and hence a finite effective (or fake) temperature. The burning question is how this universal hallmark of UV-completeness is captured in effective models of quantum gravity. What causes entropy to decrease at high energies, and deviate from the black hole area law?

In our investigation of sine dilaton gravity, we found that in the exact Hamiltonian quantization of the bulk, the distinction between temperature and fake temperature followed from imposing a certain positivity constraint on the bulk geodesic length, and that this has a semi-classical interpretation as inserting a defect in the path integral. Still, this is not a fully satisfactory final answer. Indeed, what is the universal new lesson about gravity? One would like a prescription directly at the level of the path integral of sine dilaton gravity directly, not at the Hamiltonian quantization level. Our defect picture of section \ref{sect:defect} can be interpreted as initial steps in that direction.

We now suggest several concrete open problems to make progress along these lines.
\begin{enumerate}
   \item As shown in section \ref{2}, at the classical level, the q-Schwarzian QM and sine dilaton gravity model point towards a theory with ``fake'' thermodynamics. It is not clear how this would arise in the quantum theory. Indeed, when solving the quantum dynamics of DSSYK (see section \ref{sect:background} and \cite{Berkooz:2018jqr,Berkooz:2018qkz,Blommaert:2023opb}), it seemed very natural to think of $n$ as discretized, and in particular positive $n\geq 0$. Can one also quantize without the positivity constraint, or is it a fundamental property of sine dilaton gravity? The fact that there are classical solutions (see section \ref{2}) which significantly violate $n\geq 0$ suggest that one should be able to quantize without the constraint. E.g. following \cite{Kruthoff:2024gxc} one would expect a gravity density of states with fake DSSYK thermodynamics
    \begin{equation}
        \d E\, \rho_\text{BH}(E)\sim \d \cos(\theta)\,\sinh\bigg(\frac{\pi\theta}{\abs{\log q}}\bigg)\,.
    \end{equation}
    Summing over horizons naively generates an infinite prefactor.
   
   \item What makes sine dilaton gravity special? Naively, one thing that stands out is the periodicity of the dilaton potential. This gives rise to an infinite set of horizons, which all have to be counted as independent gravitational saddles \cite{Kruthoff:2024gxc}. The resulting sum could diverge, perhaps this is a sign that a modification is needed to make sense of these models, and a sign that they are universally dual to theories with finite-support spectra. A more basic guess would be that it is the appearance of a black hole and a cosmological horizon that could signal a duality to a UV-complete (truncated spectrum) theory. To make progress on this, one could try canonical quantization of more general dilaton gravity models, along the lines of section \ref{sect:harlowjafferissine}. This seems a good idea if one wants to test claims about universality. We have made progress on canonical quantization via a Weyl-rescaled $L$ variable, and will report on this in future work. 
    
    One case for which this immediately works in an obvious way is sinh-dilaton gravity studied in \cite{Mertens:2020hbs,Fan:2021bwt,Blommaert:2023wad,Turiaci:2020fjj,Mertens:2022aou,Collier:2023cyw}. Quantization along the lines of section \ref{sect:harlowjafferissine} leads directly to the q-Schwarzian model with $\vert q \vert =1$ that the theory was proposed to be dual to in \cite{Blommaert:2023wad}, and reproduces Liouville gravity observables on the nose.

    \item Is our dilaton gravity description identical to quantum gravity on the hyperbolic quantum disk \cite{Berkooz:2023cqc,Almheiri:2024ayc}? Perhaps the boundary correlators of \cite{Almheiri:2024ayc} are to be inserted in a q-Schwarzian path integral (coupling to a type of wiggly boundary). This would be a new type of holographic duality between non-commutative AdS and a q-deformation of CFT. 
    
    In fact, one concrete result along these lines is the following. If one studies the boundary two-point function in sinh dilaton gravity \cite{Mertens:2020hbs}, in the regime where both energy integrals in the expression are dominated by large values (with a small energy difference between them), one can show that the resulting correlator reduces to precisely the semi-classical one that is governed by q-conformal symmetry \cite{Almheiri:2024ayc}. We will present this computation elsewhere \cite{toap}.
   
    \item An exact description of defects and branes in sine dilaton gravity might point to a precise gravitational path integral prescription that reproduces the partition function \eqref{4.34 magic}. Ideally, this would involve understanding the wavefunction on the defect from a physical point of view. Defect wavefunctions have many potentially interesting interpretations associated with branes.\footnote{See for instance \cite{Gao:2021uro,Blommaert:2022ucs,Okuyama:2021eju,Saad:2021uzi,Okuyama:2023byh}.} Perhaps one could be guided by representation theory of SU$_q(1,1)$ to pinpoint the precise building blocks as characters, as has been done for JT in the past \cite{Mertens:2019tcm,Belaey:2023jtr}.
     
    \item The very recent work \cite{Almheiri:2024xtw} studies the high temperature regime of DSSYK, close to $\theta=\pi/2$. In particular, they write down an effective dilaton gravity model that captures the limiting thermodynamics. Their dilaton potential depends on temperature and, as they emphasize, as a result their model is non-local. This is quite different from our work: we propose a \emph{local} dilaton gravity dual of DSSYK. It would be interesting to understand in more detail whether one could transfer between these models in an insightful way.
\end{enumerate}
\subsection{Towards observers in dS space from DSSYK?}\label{sect:5.2}

It has been claimed that DSSYK is a microscopic model of de Sitter space \cite{Susskind:2021esx,Susskind:2022bia,Lin:2022nss,Rahman:2022jsf,Susskind:2022dfz,Verlinde:2024znh,Narovlansky:2023lfz,Verlinde:2024zrh}.

We argued in section \ref{2.1} that the gravitational contour for our model follows a suitable complexified trajectory. To make contact with de Sitter physics from our model of sine dilaton gravity, one can for instance choose the gravitational contour in our case as:
\begin{equation}
    \begin{tikzpicture}[baseline={([yshift=-.5ex]current bounding box.center)}, scale=0.7]
 \pgftext{\includegraphics[scale=1]{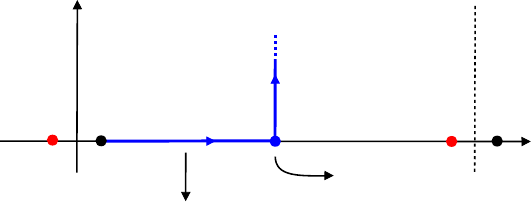}} at (0,0);
    \draw (2.1,-1.3) node {pode};
    \draw (2.1,-2) node {geodesic};
    \draw (4.6,-1.8) node {$r$};
    \draw (-1.3,-2.2) node {analogue of};
    \draw (-1.3,-2.9) node {static patch?};
    \draw (0.3,2.4) node {\color{blue}Centaur-like};
    \draw (0.3,1.7) node {\color{blue}contour?};
  \end{tikzpicture}
\end{equation}
Along this contour, $F(r)$ is real and positive. The ``turning point'' (at $r=\pi$) is a geodesic ($K=0$). If one would interpret the region between the horizon and the cosmological horizon as analogous to the static patch in dS, then this geodesic $r=\pi$ would be the worldline of the static patch observer (also called a pode). In this sense, these spacetimes are somewhat akin to the ``Centaur'' geometries used to study dS quantum gravity \cite{Anninos:2017hhn,Anninos:2020cwo} - with the difference that we have a microscopic description (namely DSSYK) from which these geometries follow. To strengthen this analogy, we note that classically the Ricci-scalar $R$ equals
\begin{equation}
R = - 2 \cos( 2 \abs{\log q} \Phi)\,.
\end{equation}
So the pode is in a regions with positive cosmological constant $R=2$. For suitably chosen $\theta\sim \pi$ the whole region between the horizons would have positive curvature:\footnote{The horizons for $\theta\sim \pi$ would be very close to the center of this picture.}
\begin{equation}
    \begin{tikzpicture}[baseline={([yshift=-.5ex]current bounding box.center)}, scale=0.7]
 \pgftext{\includegraphics[scale=1]{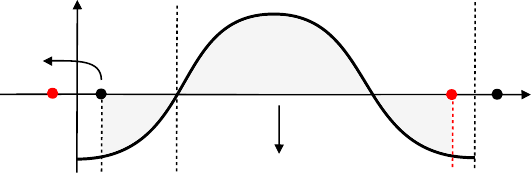}} at (0,0);
    \draw (-4.9,0.5) node {horizon};
    \draw (0.7,-1.6) node {positive};
    \draw (0.7,-2.3) node {curvature};
    \draw (4.4,-1) node {$r$};
    \draw (3.5,-1.8) node {$2\pi$};
    \draw (-1.5,-1.8) node {$\pi/2$};
    \draw (-3.25,-1.8) node {$0$};
    \draw (-2.75,-1.8) node {$\theta$};
    \draw (1.4,1.4) node {$R$};
  \end{tikzpicture}
\end{equation}

Much like DSSYK, dS space too has a distinction between temperature and fake temperature. In our case, as we have emphasized, this distinction follows from the length positivity constraint \eqref{1.7 constraint}. If one takes the relation of DSSYK with dS seriously, the logical conclusion is that our length constraint \eqref{1.7 constraint} should explain in dS why there are these two notions of temperature. Such an explanation already exists to some degree. It seems related with the fact that in cosmology the observer should be included in the quantum description of the system \cite{Chandrasekaran:2022cip,Witten:2023xze}. This suggests that the fundamental lesson to learn here, is what the precise quantum mechanical description of an observer in dS space is (at least in our simple toy model). 

Next we discuss a different way to make contact with dS physics.

We will search a possible interpretation for the defect wavefunction of section \ref{sect:defect}, along the lines of the above discussion. In particular, in the spirit of \eqref{4.39 fixeda} we ask for the interpretation of the additional wavefunction
\begin{equation}
    \psi(A)=e^{-4\abs{\log q} A^2}\,,\quad A=\frac{\theta}{2\abs{\log q}}
\end{equation}
in the sine dilaton gravity path integral. This represents the $\theta^2$ term in the entropy of DSSYK \eqref{3.1 ent} which (as we emphasized ad nauseam in this work) is responsible for the difference between temperature and fake temperature, and essentially encodes the constraint \eqref{1.7 constraint}. We have not yet understood this in general. But close to the maximum of the entropy $\theta=\pi/2$ we remark that it matches with the substitution \cite{Witten:2023xze}
\begin{equation}
    e^{S_\text{BH}}=e^{2\pi A} \,\to\, e^S=e^{2\pi A}e^{- m B}\,.
\end{equation}
Here $m$ is the mass of a bulk particle which we constrain (quite ad-hoc) to be equal to the ADM energy \eqref{2.bb energy} of the spacetime. We think of this as analogous to the energy constraints in closed spacetimes. $B$ denotes the length of a closed geodesic followed by this non-minimally coupled particle:
\begin{equation}
    B=\oint\d s\,e^{-\i\abs{\log q}\Phi}\,.
\end{equation}
The geodesic $B$ in question that we propose is the following. The region in between the ``interior'' and outer horizon (see equation \eqref{2.39 pic}) of the effective AdS$_2$ metric \eqref{eff} is minus a Euclidean two sphere. Indeed, in this region $-\sin(\theta)<\rho<\sin(\theta)$ one can use the coordinates
\begin{equation}
    \rho=\cos(\chi)\sin(\theta)\,,\quad 0<\chi<\pi\,,
\end{equation}
resulting in the Euclidean sphere metric
\begin{equation}
    -\d s^2_\text{eff}=\d \chi^2+\sin(\theta)^2\sin(\chi)^2\d \tau^2\,.
\end{equation}
This sphere has a closed geodesic at $\chi=\pi/2$ which one would call the pode or static observer worldline, when interpreting this sphere as (minus) Euclidean dS space. We consider $B$ to be this worldline. Now, if we consider geometries very close to the entropy maximum one obtains
\begin{equation}
    \theta=\frac{\pi}{2}+\epsilon\,,\quad m=-\frac{\cos(\theta)}{2\abs{\log q}}\approx \frac{\epsilon}{2\abs{\log q}}\,,\quad B= 2 \pi \sin(\theta) \approx 2 \pi\,,\quad \frac{\theta^2}{\abs{\log q}}\approx \text{constant} + \frac{2 \pi \epsilon}{2\abs{\log q}}\,.
\end{equation}
The observation is that in this regime indeed adding the term $m B$ matches with the difference between DSSYK and the naive sine dilaton gravity entropies \eqref{3.1 ent} respectively \eqref{2.bb energy}
\begin{equation}
    \frac{\theta^2}{\abs{\log q}}+\text{constant}=m B\,.
\end{equation}

The substitution in \cite{Witten:2023xze} was motivated by actually including a physical observer in dS gravity. It seems that DSSYK is implementing this procedure, at least in the regime where one expects to capture small fluctuations away from empty dS. Of course this argument was schematic. One would want to understand an exact quantum mechanical implementation of such an idea, and obtain an all-energies match with DSSYK.

\section*{Acknowledgments}
We thank Micha Berkooz, Alejandro Cabo-Bizet, Damian Galante, Victor Gorbenko, Mikhail Isachenkov, Jorrit Kruthoff, Adam Levine, Ohad Mamroud, Vladimir Narovlansky, Thomas Tappeiner, Herman Verlinde, and especially Shunyu Yao for useful discussions. AB was supported by the ERC-COG Grant NP-QFT No. 864583 and by INFN Iniziativa Specifica GAST. TM and JP acknowledge financial support from the European Research Council (grant BHHQG-101040024). Funded by the European Union. Views and opinions expressed are however those of the author(s) only and do not necessarily reflect those of the European Union or the European Research Council. Neither the European Union nor the granting authority can be held responsible for them. 

\appendix

\section{From DSSYK amplitudes towards the Feynman path integral}
\label{app:fpi}
The classical Hamiltonian of \eqref{1.10 hamil} is $2\pi$-periodic in $p$. This makes this dynamical system analogous to that of a particle on a periodic space, with a swapping of $x \leftrightarrow p$.
Consider hence the Feynman path integral for a particle on a circle $S^1$ with $x \sim x+2\pi$:
\begin{align}
\langle x_f,t_f \vert x_i,t_i \rangle = \prod_{n=1}^{N}\left[\int_0^{2\pi}dx_n\right] \prod_{n=1}^{N+1}\left[\sum_{p_n/\hbar=-\infty}^{+\infty}\right] e^{i \sum_{n=1}^{N+1}\left[\frac{p_n}{\hbar}(x_n-x_{n-1}) - \frac{\epsilon}{\hbar}H(p_n,x_n)\right]}
\end{align}
where we inserted complete sets of states in both position and momentum basis as usual. The Hamiltonian $H$ is an arbitrary function on the circle at this point. This form has no direct continuum limit, since the momenta are discrete with separation $\hbar$. We can alternatively rewrite this expression by adding a summation over winding numbers in the universal covering space as (see e.g. \cite{Kleinert:788154})
\begin{align}
\langle x_f,t_f \vert x_i,t_i \rangle &= \sum_{l=-\infty}^{+\infty} \langle x_f +2 \pi l,t_f \vert x_i,t_i \rangle_{\text{non-cyclic}} \\
&=\sum_{l=-\infty}^{+\infty}\prod_{n=1}^{N}\left[\int_{-\infty}^{+\infty}dx_n\right] \prod_{n=1}^{N+1} \left[\int_{-\infty}^{+\infty}dp_n\right] e^{i \sum_{n=1}^{N+1}\left[\frac{p_n}{\hbar}(x_n-x_{n-1}) + \frac{p_{N+1}}{\hbar}2\pi l - \frac{\epsilon}{\hbar}H(p_n,x_n)\right]}
\end{align}
which now has a direct continuum limit. Of more interest in our case, is the ``conjugate'' evolution amplitude:
\begin{equation}
\langle p_f,t_f \vert p_i,t_i \rangle = \int_0^{2\pi} dx_i dx_f e^{i x_i \frac{p_i}{\hbar} - i x_f \frac{p_f}{\hbar}} \langle x_f,t_f \vert x_i,t_i \rangle
\end{equation}
The integration ranges of $x_i$ and $x_f$ can be extended to $(-\infty,+\infty)$ by absorbing the summation over $l$ into one of them (because the bra-ket on the right is a periodic function of position). We end up with the continuum description:
\begin{equation}
\langle p_f,t_f \vert p_i,t_i \rangle = \frac{1}{\sum_{l=-\infty}^{+\infty} 1}\int_{p_i}^{p_f} \mathcal{D} x\mathcal{D} p \, e^{ iS[q,p]}
\end{equation}
The divergent factor in the denominator $\sum_{l=-\infty}^{+\infty} 1 = V_{\text{gauge}}$ compensates for the option of rigidly moving any given path $(x(t),p(t))$ to $(x(t)+2\pi l,p(t))$ with the same weight in the path integral, and is interpreted as the volume of the gauged symmetry group. We emphasize that the initial and final $p_i,p_f$ have to be discrete $\in \hbar\mathbb{Z}$, but intermediate paths do not. 
Finally reinterpreting $x \leftrightarrow p$ and plugging in the $2\pi$-periodic DSSYK Hamiltonian \eqref{1.10 hamil}, we arrive at the phase space path integral description of \eqref{1.20 bis}.

\section{Boundary conditions and the Poisson-sigma model interpretation}
\label{s:PSM}
Of relevance to our discussion will be the fact that the theory \eqref{1.20 bis} is obtained from a more general theory (particle on SU$_q(1,1)$) with 6 fields $(\varphi,\beta,\gamma,p_\varphi,p_\beta,p_\gamma)$, by imposing two constraints, one of which reads
\begin{equation}
    e^{-\i\abs{\log q}h} f=\frac{\i}{2\abs{\log q}}\,.\label{1.21 const}
\end{equation}
Here $f$ and $h$ are explicit functions of $(\varphi,\beta,\gamma,p_\varphi,p_\beta,p_\gamma)$ given in equation (4.8) of \cite{Blommaert:2023opb} that are (together with another function $e$) generators of an SU$_q(1,1)$ algebra:
\begin{equation}
    \{h,e\}=e\,,\quad  \{h,f\}=-f\,,\quad  \{e,f\}=\frac{\sin(2\abs{\log q}h)}{\abs{\log q}}\,.\label{chargealgebra}
\end{equation}
The constraints \eqref{1.21 const} have an important interpretation as implementing Brown-Henneaux-type boundary conditions on sine dilaton gravity, as we discuss in section \ref{2.1}.

Along the lines of section 3.3 in \cite{Blommaert:2023wad}, the boundary conditions follow from translating the constraints \eqref{1.21 const} that have to be imposed on quantum mechanics on a quantum group SU$_q(1,1)$ in order to reduce it to the q-Schwarzian quantum mechanical model \eqref{1.20 bis}. Via the Poisson-sigma model approach \cite{Blommaert:2023wad} one deduces $f=\sqrt{h}$ and $h=-\Phi_\text{bdy}$. This indeed reduces \eqref{1.21 const} to the gravitational boundary conditions \eqref{2.3 bc}.

\section{q-Schwarzian entropy with constraints}\label{A1}
Here we compute the integral \eqref{I4}. It is convenient to split the integral in \eqref{g(z)} in the following way
\begin{equation}\label{I}
\int_{\gamma_2}\d z\, g(z)=\int_{e^{-4i\theta}}^{1} \mathrm{d}z  \left(\frac{1}{2z}+\frac{e^{2 i \theta}}{1-e^{2i \theta}z}\right)\left(-\i\theta+\log\left(1-e^{4\i\theta}z\right)-\log(1-e^{2\i\theta}z)\right)\,.
\end{equation} 
Now, every integral appearing in \eqref{I} can be performed separately by doing some manipulations and exploiting the following integral representation of the dilogarithm
\begin{equation}
\mathrm{Li}_{2}(z)=-\int_{0}^{z} \mathrm{d}u \, \frac{\log(1-u)}{u}\,.
\end{equation}
This results in
\begin{align}
\int_{\gamma_2}\d z\, g(z)&=\pi \theta +\frac{\pi^2}{12}+\frac12 \mathrm{Li}_{2}\left(e^{2\i\theta}\right)-\frac12\mathrm{Li}_{2}\left(e^{-2\i\theta}\right)-\frac12 \mathrm{Li}_{2}\left(e^{4\i\theta}\right)\nonumber\\ 
&\qquad -\log \left(-e^{2 \i \theta }\right) \log \left(1-e^{4 \i \theta }\right)-\text{Li}_2\left(1+e^{2 \i \theta }\right)+\frac12 \log^2\left(1-e^{2 \i\theta}\right)-\frac12 \log^2\left(1-e^{-2\i\theta}\right)\nonumber\\
&=\pi \theta -\frac{\pi^2}{12}-\frac12 \mathrm{Li}_{2}\left(e^{2\i\theta}\right)-\frac12\mathrm{Li}_{2}\left(e^{-2\i\theta}\right)+\log \left(-e^{2 \i \theta }\right) \log \left(1+e^{2 \i \theta }\right) \nonumber\\
&\qquad -\log \left(-e^{2 \i \theta }\right) \log \left(1-e^{4 \i \theta }\right)+\frac12 \log^2\left(1-e^{2 \i\theta}\right)-\frac12 \log^2\left(1-e^{-2\i\theta}\right)\,.
\end{align}
In the second equality we used the following two dilogarithm identities
\begin{equation}\label{dilog_ide}
\text{Li}_2\left(x\right)+\text{Li}_2\left(1-x\right)=\frac{\pi^2}{6}-\log(x)\log(1-x)\,,\quad \text{Li}_2\left(x\right)+\text{Li}_2\left(-x\right)=\frac12\text{Li}_2\left(x^2\right)\,.
\end{equation} 
The logarithms that remain can be simplified to obtain equation \eqref{I4}.  
\section{DSSYK semi-classical four-point function}\label{app:DSSYK4pt}
We consider an ``uncrossed'' four-point function with two horizontal lines. We label energies by (bottom to top) $\theta_1,\theta_2,\theta_3$, assign thermal weights (bottom to top) $\tau_1,\tau_2,\beta-\tau_1-\tau_2$, and consider non-identical operator dimensions $\Delta_1,\Delta_2$. The exact DSSYK four-point function with these specifications is
\begin{align}
    \frac{1}{Z(\beta)}\int_0^\pi \d\theta_1\,\rho(\theta_1)\int_0^\pi \d\theta_2\,\rho(\theta_2)\int_0^\pi \d\theta_3\,\rho(\theta_3)&\exp\bigg((\beta-\tau_1-\tau_2)\frac{\cos(\theta_3)}{2\abs{\log q}}+\tau_1 \frac{\cos(\theta_1)}{2\abs{\log q}}+\tau_2 \frac{\cos(\theta_2)}{2\abs{\log q}}\bigg)\nonumber\\&\frac{(q^{4\Delta_1};q^2)_\infty}{(q^{2\Delta_1}e^{\pm\i\theta_1\pm\i\theta_2};q^2)_\infty}\,\frac{(q^{4\Delta_2};q^2)_\infty}{(q^{2\Delta_2}e^{\pm\i\theta_1\pm\i\theta_3};q^2)_\infty}\,.
\end{align}
We consider the following parametrization of energies
\begin{align}
    \theta_1&=\theta-\a_1\abs{\log q}-\a_2\abs{\log q}\nonumber\\
    \theta_2&=\theta+\a_1\abs{\log q}-\a_2\abs{\log q}\nonumber\\
    \theta_1&=\theta+\a_1\abs{\log q}+\a_2\abs{\log q}\,.
\end{align}
In the $\abs{\log q}\to 0$ semi-classical limit one then finds
\begin{align}
    &\frac{1}{Z(\beta)}\int_0^\pi\d\cos(\theta)\,\exp\bigg( \frac{\pi\theta-\theta^2}{\abs{\log q}} +\beta \frac{\cos(\theta)}{2\abs{\log q}}\bigg)\label{b.3 int}\\&
    \sin(\theta)^{2\Delta_1}\int_{-\infty}^{+\infty}\d\a_1\,e^{(\tau_1-\beta/2)\a_1\sin(\theta)}\frac{\Gamma(\Delta_1\pm\i\a_1)}{\Gamma(2\Delta_1)}\,\sin(\theta)^{2\Delta_2}\int_{-\infty}^{+\infty}\d\a_2\,e^{(\tau_2-\beta/2+\tau_1)\a_2\sin(\theta)}\frac{\Gamma(\Delta_2\pm\i\a_2)}{\Gamma(2\Delta_2)}\,.\nonumber
\end{align}
We are interested in two-sided correlators where both the particle lines cross the ERB, one at Lorentzian time $T_1$, and one at Lorentzian time $T_2$. The correct parametrization of Euclidean times for this is
\begin{equation}
    \tau_1=\beta/2+\i T_1\,,\quad \tau_2=\i(T_2-\i T_1)\,.
\end{equation}
Crucially the integrals on the second line of \eqref{b.3 int} become independent of $\beta$ and evaluate for every fixed $\theta$ to
\begin{equation}
    \frac{\sin(\theta)^{2\Delta_1}}{\cosh(\sin(\theta) T_1/2)^{2\Delta_1}}\,\frac{\sin(\theta)^{2\Delta_2}}{\cosh(\sin(\theta) T_2/2)^{2\Delta_2}}\,.\label{2.5}
\end{equation}
This reproduces the sine dilaton gravity prediction \eqref{4.44 pic}.


\bibliographystyle{utphys}
\bibliography{Refs}

\providecommand{\href}[2]{#2}\begingroup\raggedright\begin{thebibliography}{10}

\bibitem{Blommaert:2023opb}
A.~Blommaert, T.~G. Mertens, and S.~Yao, ``{Dynamical actions and q-representation theory for double-scaled SYK},'' \href{http://arxiv.org/abs/2306.00941}{{\ttfamily arXiv:2306.00941 [hep-th]}}.

\bibitem{Blommaert:2023wad}
A.~Blommaert, T.~G. Mertens, and S.~Yao, ``{The q-Schwarzian and Liouville gravity},'' \href{http://arxiv.org/abs/2312.00871}{{\ttfamily arXiv:2312.00871 [hep-th]}}.

\bibitem{Lin:2022rbf}
H.~W. Lin, ``{The bulk Hilbert space of double scaled SYK},'' \href{http://dx.doi.org/10.1007/JHEP11(2022)060}{{\em JHEP} {\bfseries 11} (2022) 060}, \href{http://arxiv.org/abs/2208.07032}{{\ttfamily arXiv:2208.07032 [hep-th]}}.

\bibitem{kitaev2015simple}
A.~Kitaev, ``A simple model of quantum holography (part 2),'' {\em Entanglement in Strongly-Correlated Quantum Matter} (2015) 38.

\bibitem{Maldacena:2016hyu}
J.~Maldacena and D.~Stanford, ``{Remarks on the Sachdev-Ye-Kitaev model},'' \href{http://dx.doi.org/10.1103/PhysRevD.94.106002}{{\em Phys. Rev. D} {\bfseries 94} no.~10, (2016) 106002}, \href{http://arxiv.org/abs/1604.07818}{{\ttfamily arXiv:1604.07818 [hep-th]}}.

\bibitem{Maldacena:2016upp}
J.~Maldacena, D.~Stanford, and Z.~Yang, ``{Conformal symmetry and its breaking in two dimensional Nearly Anti-de-Sitter space},'' \href{http://dx.doi.org/10.1093/ptep/ptw124}{{\em PTEP} {\bfseries 2016} no.~12, (2016) 12C104}, \href{http://arxiv.org/abs/1606.01857}{{\ttfamily arXiv:1606.01857 [hep-th]}}.

\bibitem{Engelsoy:2016xyb}
J.~Engels\"oy, T.~G. Mertens, and H.~Verlinde, ``{An investigation of AdS$_{2}$ backreaction and holography},'' \href{http://dx.doi.org/10.1007/JHEP07(2016)139}{{\em JHEP} {\bfseries 07} (2016) 139}, \href{http://arxiv.org/abs/1606.03438}{{\ttfamily arXiv:1606.03438 [hep-th]}}.

\bibitem{Jensen:2016pah}
K.~Jensen, ``{Chaos in AdS$_2$ Holography},'' \href{http://dx.doi.org/10.1103/PhysRevLett.117.111601}{{\em Phys. Rev. Lett.} {\bfseries 117} no.~11, (2016) 111601}, \href{http://arxiv.org/abs/1605.06098}{{\ttfamily arXiv:1605.06098 [hep-th]}}.

\bibitem{Mertens:2022irh}
T.~G. Mertens and G.~J. Turiaci, ``{Solvable Models of Quantum Black Holes: A Review on Jackiw-Teitelboim Gravity},'' \href{http://arxiv.org/abs/2210.10846}{{\ttfamily arXiv:2210.10846 [hep-th]}}.

\bibitem{Cotler:2016fpe}
J.~S. Cotler, G.~Gur-Ari, M.~Hanada, J.~Polchinski, P.~Saad, S.~H. Shenker, D.~Stanford, A.~Streicher, and M.~Tezuka, ``{Black Holes and Random Matrices},'' \href{http://dx.doi.org/10.1007/JHEP05(2017)118}{{\em JHEP} {\bfseries 05} (2017) 118}, \href{http://arxiv.org/abs/1611.04650}{{\ttfamily arXiv:1611.04650 [hep-th]}}. [Erratum: JHEP 09, 002 (2018)].

\bibitem{Berkooz:2018jqr}
M.~Berkooz, M.~Isachenkov, V.~Narovlansky, and G.~Torrents, ``{Towards a full solution of the large N double-scaled SYK model},'' \href{http://dx.doi.org/10.1007/JHEP03(2019)079}{{\em JHEP} {\bfseries 03} (2019) 079}, \href{http://arxiv.org/abs/1811.02584}{{\ttfamily arXiv:1811.02584 [hep-th]}}.

\bibitem{Berkooz:2018qkz}
M.~Berkooz, P.~Narayan, and J.~Simon, ``{Chord diagrams, exact correlators in spin glasses and black hole bulk reconstruction},'' \href{http://dx.doi.org/10.1007/JHEP08(2018)192}{{\em JHEP} {\bfseries 08} (2018) 192}, \href{http://arxiv.org/abs/1806.04380}{{\ttfamily arXiv:1806.04380 [hep-th]}}.

\bibitem{Jafferis:2022wez}
D.~L. Jafferis, D.~K. Kolchmeyer, B.~Mukhametzhanov, and J.~Sonner, ``{JT gravity with matter, generalized ETH, and Random Matrices},'' \href{http://arxiv.org/abs/2209.02131}{{\ttfamily arXiv:2209.02131 [hep-th]}}.

\bibitem{Susskind:2022bia}
L.~Susskind, ``{De Sitter Space, Double-Scaled SYK, and the Separation of Scales in the Semiclassical Limit},'' \href{http://arxiv.org/abs/2209.09999}{{\ttfamily arXiv:2209.09999 [hep-th]}}.

\bibitem{Bhattacharjee:2022ave}
B.~Bhattacharjee, P.~Nandy, and T.~Pathak, ``{Krylov complexity in large-$q$ and double-scaled SYK model},'' \href{http://arxiv.org/abs/2210.02474}{{\ttfamily arXiv:2210.02474 [hep-th]}}.

\bibitem{Susskind:2023hnj}
L.~Susskind, ``{De Sitter Space has no Chords. Almost Everything is Confined},'' \href{http://arxiv.org/abs/2303.00792}{{\ttfamily arXiv:2303.00792 [hep-th]}}.

\bibitem{Mukhametzhanov:2023tcg}
B.~Mukhametzhanov, ``{Large p SYK from chord diagrams},'' \href{http://arxiv.org/abs/2303.03474}{{\ttfamily arXiv:2303.03474 [hep-th]}}.

\bibitem{Berkooz:2023cqc}
M.~Berkooz, Y.~Jia, and N.~Silberstein, ``{Parisi's hypercube, Fock-space frustration and near-AdS$_2$/near-CFT$_1$ holography},'' \href{http://arxiv.org/abs/2303.18182}{{\ttfamily arXiv:2303.18182 [hep-th]}}.

\bibitem{Okuyama:2023bch}
K.~Okuyama and K.~Suzuki, ``{Correlators of double scaled SYK at one-loop},'' \href{http://arxiv.org/abs/2303.07552}{{\ttfamily arXiv:2303.07552 [hep-th]}}.

\bibitem{Lin:2022nss}
H.~Lin and L.~Susskind, ``{Infinite Temperature's Not So Hot},'' \href{http://arxiv.org/abs/2206.01083}{{\ttfamily arXiv:2206.01083 [hep-th]}}.

\bibitem{Berkooz:2022mfk}
M.~Berkooz, M.~Isachenkov, P.~Narayan, and V.~Narovlansky, ``{Quantum groups, non-commutative $AdS_2$, and chords in the double-scaled SYK model},'' \href{http://arxiv.org/abs/2212.13668}{{\ttfamily arXiv:2212.13668 [hep-th]}}.

\bibitem{Goel:2023svz}
A.~Goel, V.~Narovlansky, and H.~Verlinde, ``{Semiclassical geometry in double-scaled SYK},'' \href{http://arxiv.org/abs/2301.05732}{{\ttfamily arXiv:2301.05732 [hep-th]}}.

\bibitem{Narovlansky:2023lfz}
V.~Narovlansky and H.~Verlinde, ``{Double-scaled SYK and de Sitter Holography},'' \href{http://arxiv.org/abs/2310.16994}{{\ttfamily arXiv:2310.16994 [hep-th]}}.

\bibitem{Verlinde:2024zrh}
H.~Verlinde and M.~Zhang, ``{SYK Correlators from 2D Liouville-de Sitter Gravity},'' \href{http://arxiv.org/abs/2402.02584}{{\ttfamily arXiv:2402.02584 [hep-th]}}.

\bibitem{Berkooz:2024evs}
M.~Berkooz, N.~Brukner, Y.~Jia, and O.~Mamroud, ``{From Chaos to Integrability in Double Scaled SYK},'' \href{http://arxiv.org/abs/2403.01950}{{\ttfamily arXiv:2403.01950 [hep-th]}}.

\bibitem{Lin:2023trc}
H.~W. Lin and D.~Stanford, ``{A symmetry algebra in double-scaled SYK},'' \href{http://arxiv.org/abs/2307.15725}{{\ttfamily arXiv:2307.15725 [hep-th]}}.

\bibitem{Verlinde:2024znh}
H.~Verlinde, ``{Double-scaled SYK, Chords and de Sitter Gravity},'' \href{http://arxiv.org/abs/2402.00635}{{\ttfamily arXiv:2402.00635 [hep-th]}}.

\bibitem{Almheiri:2024ayc}
A.~Almheiri and F.~K. Popov, ``{Holography on the Quantum Disk},'' \href{http://arxiv.org/abs/2401.05575}{{\ttfamily arXiv:2401.05575 [hep-th]}}.

\bibitem{Almheiri:2024xtw}
A.~Almheiri, A.~Goel, and X.-Y. Hu, ``{Quantum gravity of the Heisenberg algebra},'' \href{http://arxiv.org/abs/2403.18333}{{\ttfamily arXiv:2403.18333 [hep-th]}}.

\bibitem{hawking1975particle}
S.~W. Hawking, ``Particle creation by black holes,'' {\em Communications in mathematical physics} {\bfseries 43} no.~3, (1975) 199--220.

\bibitem{Streicher:2019wek}
A.~Streicher, ``{SYK Correlators for All Energies},'' \href{http://dx.doi.org/10.1007/JHEP02(2020)048}{{\em JHEP} {\bfseries 02} (2020) 048}, \href{http://arxiv.org/abs/1911.10171}{{\ttfamily arXiv:1911.10171 [hep-th]}}.

\bibitem{Choi:2019bmd}
C.~Choi, M.~Mezei, and G.~S\'arosi, ``{Exact four point function for large $q$ SYK from Regge theory},'' \href{http://dx.doi.org/10.1007/JHEP05(2021)166}{{\em JHEP} {\bfseries 05} (2021) 166}, \href{http://arxiv.org/abs/1912.00004}{{\ttfamily arXiv:1912.00004 [hep-th]}}.

\bibitem{Dong:2022ilf}
X.~Dong, D.~Marolf, P.~Rath, A.~Tajdini, and Z.~Wang, ``{The spacetime geometry of fixed-area states in gravitational systems},'' \href{http://dx.doi.org/10.1007/JHEP08(2022)158}{{\em JHEP} {\bfseries 08} (2022) 158}, \href{http://arxiv.org/abs/2203.04973}{{\ttfamily arXiv:2203.04973 [hep-th]}}.

\bibitem{Chandrasekaran:2022cip}
V.~Chandrasekaran, R.~Longo, G.~Penington, and E.~Witten, ``{An algebra of observables for de Sitter space},'' \href{http://dx.doi.org/10.1007/JHEP02(2023)082}{{\em JHEP} {\bfseries 02} (2023) 082}, \href{http://arxiv.org/abs/2206.10780}{{\ttfamily arXiv:2206.10780 [hep-th]}}.

\bibitem{Witten:2023xze}
E.~Witten, ``{A Background Independent Algebra in Quantum Gravity},'' \href{http://arxiv.org/abs/2308.03663}{{\ttfamily arXiv:2308.03663 [hep-th]}}.

\bibitem{Susskind:2021esx}
L.~Susskind, ``{Entanglement and Chaos in De Sitter Space Holography: An SYK Example},'' \href{http://dx.doi.org/10.22128/jhap.2021.455.1005}{{\em JHAP} {\bfseries 1} no.~1, (2021) 1--22}, \href{http://arxiv.org/abs/2109.14104}{{\ttfamily arXiv:2109.14104 [hep-th]}}.

\bibitem{Rahman:2022jsf}
A.~A. Rahman, ``{dS JT Gravity and Double-Scaled SYK},'' \href{http://arxiv.org/abs/2209.09997}{{\ttfamily arXiv:2209.09997 [hep-th]}}.

\bibitem{Susskind:2022dfz}
L.~Susskind, ``{Scrambling in Double-Scaled SYK and De Sitter Space},'' \href{http://arxiv.org/abs/2205.00315}{{\ttfamily arXiv:2205.00315 [hep-th]}}.

\bibitem{Harlow:2018tqv}
D.~Harlow and D.~Jafferis, ``{The Factorization Problem in Jackiw-Teitelboim Gravity},'' \href{http://dx.doi.org/10.1007/JHEP02(2020)177}{{\em JHEP} {\bfseries 02} (2020) 177}, \href{http://arxiv.org/abs/1804.01081}{{\ttfamily arXiv:1804.01081 [hep-th]}}.

\bibitem{Gegenberg:1994pv}
J.~Gegenberg, G.~Kunstatter, and D.~Louis-Martinez, ``{Observables for two-dimensional black holes},'' \href{http://dx.doi.org/10.1103/PhysRevD.51.1781}{{\em Phys. Rev. D} {\bfseries 51} (1995) 1781--1786}, \href{http://arxiv.org/abs/gr-qc/9408015}{{\ttfamily arXiv:gr-qc/9408015}}.

\bibitem{Witten:2020ert}
E.~Witten, ``{Deformations of JT Gravity and Phase Transitions},'' \href{http://arxiv.org/abs/2006.03494}{{\ttfamily arXiv:2006.03494 [hep-th]}}.

\bibitem{brown1993quasilocal}
J.~D. Brown and J.~W. York~Jr, ``Quasilocal energy and conserved charges derived from the gravitational action,'' {\em Physical Review D} {\bfseries 47} no.~4, (1993) 1407.

\bibitem{Kruthoff:2024gxc}
J.~Kruthoff and A.~Levine, ``{Semi-classical dilaton gravity and the very blunt defect expansion},'' \href{http://arxiv.org/abs/2402.10162}{{\ttfamily arXiv:2402.10162 [hep-th]}}.

\bibitem{McGough:2016lol}
L.~McGough, M.~Mezei, and H.~Verlinde, ``{Moving the CFT into the bulk with $ T\overline{T} $},'' \href{http://dx.doi.org/10.1007/JHEP04(2018)010}{{\em JHEP} {\bfseries 04} (2018) 010}, \href{http://arxiv.org/abs/1611.03470}{{\ttfamily arXiv:1611.03470 [hep-th]}}.

\bibitem{Gross:2019ach}
D.~J. Gross, J.~Kruthoff, A.~Rolph, and E.~Shaghoulian, ``{$T\overline{T}$ in AdS$_2$ and Quantum Mechanics},'' \href{http://dx.doi.org/10.1103/PhysRevD.101.026011}{{\em Phys. Rev. D} {\bfseries 101} no.~2, (2020) 026011}, \href{http://arxiv.org/abs/1907.04873}{{\ttfamily arXiv:1907.04873 [hep-th]}}.

\bibitem{Iliesiu:2020zld}
L.~V. Iliesiu, J.~Kruthoff, G.~J. Turiaci, and H.~Verlinde, ``{JT gravity at finite cutoff},'' \href{http://dx.doi.org/10.21468/SciPostPhys.9.2.023}{{\em SciPost Phys.} {\bfseries 9} (2020) 023}, \href{http://arxiv.org/abs/2004.07242}{{\ttfamily arXiv:2004.07242 [hep-th]}}.

\bibitem{Griguolo:2021wgy}
L.~Griguolo, R.~Panerai, J.~Papalini, and D.~Seminara, ``{Nonperturbative effects and resurgence in Jackiw-Teitelboim gravity at finite cutoff},'' \href{http://dx.doi.org/10.1103/PhysRevD.105.046015}{{\em Phys. Rev. D} {\bfseries 105} no.~4, (2022) 046015}, \href{http://arxiv.org/abs/2106.01375}{{\ttfamily arXiv:2106.01375 [hep-th]}}.

\bibitem{Blommaert:2024whf}
A.~Blommaert, A.~Levine, T.~G. Mertens, J.~Papalini, and K.~Parmentier, ``{An entropic puzzle in periodic dilaton gravity and DSSYK},'' \href{http://arxiv.org/abs/2411.16922}{{\ttfamily arXiv:2411.16922 [hep-th]}}.

\bibitem{Collier:2024kmo}
S.~Collier, L.~Eberhardt, B.~M\"uhlmann, and V.~A. Rodriguez, ``{The complex Liouville string},'' \href{http://arxiv.org/abs/2409.17246}{{\ttfamily arXiv:2409.17246 [hep-th]}}.

\bibitem{yao2018edge}
S.~Yao and Z.~Wang, ``Edge states and topological invariants of non-hermitian systems,'' {\em Physical review letters} {\bfseries 121} no.~8, (2018) 086803.

\bibitem{Iliesiu:2024cnh}
L.~V. Iliesiu, A.~Levine, H.~W. Lin, H.~Maxfield, and M.~Mezei, ``{On the non-perturbative bulk Hilbert space of JT gravity},'' \href{http://arxiv.org/abs/2403.08696}{{\ttfamily arXiv:2403.08696 [hep-th]}}.

\bibitem{Saad:2018bqo}
P.~Saad, S.~H. Shenker, and D.~Stanford, ``{A semiclassical ramp in SYK and in gravity},'' \href{http://arxiv.org/abs/1806.06840}{{\ttfamily arXiv:1806.06840 [hep-th]}}.

\bibitem{Saad:2019pqd}
P.~Saad, ``{Late Time Correlation Functions, Baby Universes, and ETH in JT Gravity},'' \href{http://arxiv.org/abs/1910.10311}{{\ttfamily arXiv:1910.10311 [hep-th]}}.

\bibitem{Blommaert:2019hjr}
A.~Blommaert, T.~G. Mertens, and H.~Verschelde, ``{Clocks and rods in Jackiw-Teitelboim quantum gravity},'' \href{http://dx.doi.org/10.1007/JHEP09(2019)060}{{\em JHEP} {\bfseries 09} (2019) 060}, \href{http://arxiv.org/abs/1902.11194}{{\ttfamily arXiv:1902.11194 [hep-th]}}.

\bibitem{Almheiri:2019qdq}
A.~Almheiri, T.~Hartman, J.~Maldacena, E.~Shaghoulian, and A.~Tajdini, ``{Replica wormholes and the entropy of Hawking radiation},'' \href{http://dx.doi.org/10.1007/JHEP05(2020)013}{{\em JHEP} {\bfseries 05} (2020) 013}, \href{http://arxiv.org/abs/1911.12333}{{\ttfamily arXiv:1911.12333 [hep-th]}}.

\bibitem{Penington:2019kki}
G.~Penington, S.~H. Shenker, D.~Stanford, and Z.~Yang, ``{Replica wormholes and the black hole interior},'' \href{http://dx.doi.org/10.1007/JHEP03(2022)205}{{\em JHEP} {\bfseries 03} (2022) 205}, \href{http://arxiv.org/abs/1911.11977}{{\ttfamily arXiv:1911.11977 [hep-th]}}.

\bibitem{Iliesiu:2021ari}
L.~V. Iliesiu, M.~Mezei, and G.~S\'arosi, ``{The volume of the black hole interior at late times},'' \href{http://dx.doi.org/10.1007/JHEP07(2022)073}{{\em JHEP} {\bfseries 07} (2022) 073}, \href{http://arxiv.org/abs/2107.06286}{{\ttfamily arXiv:2107.06286 [hep-th]}}.

\bibitem{Blommaert:2020seb}
A.~Blommaert, ``{Dissecting the ensemble in JT gravity},'' \href{http://dx.doi.org/10.1007/JHEP09(2022)075}{{\em JHEP} {\bfseries 09} (2022) 075}, \href{http://arxiv.org/abs/2006.13971}{{\ttfamily arXiv:2006.13971 [hep-th]}}.

\bibitem{Blommaert:2022lbh}
A.~Blommaert, J.~Kruthoff, and S.~Yao, ``{An integrable road to a perturbative plateau},'' \href{http://dx.doi.org/10.1007/JHEP04(2023)048}{{\em JHEP} {\bfseries 04} (2023) 048}, \href{http://arxiv.org/abs/2208.13795}{{\ttfamily arXiv:2208.13795 [hep-th]}}.

\bibitem{Saad:2022kfe}
P.~Saad, D.~Stanford, Z.~Yang, and S.~Yao, ``{A convergent genus expansion for the plateau},'' \href{http://arxiv.org/abs/2210.11565}{{\ttfamily arXiv:2210.11565 [hep-th]}}.

\bibitem{Griguolo:2023jyy}
L.~Griguolo, J.~Papalini, L.~Russo, and D.~Seminara, ``{The resurgence of the plateau in supersymmetric ${N}=1$ Jackiw-Teitelboim gravity},'' \href{http://arxiv.org/abs/2310.06768}{{\ttfamily arXiv:2310.06768 [hep-th]}}.

\bibitem{Griguolo:2024htx}
L.~Griguolo, J.~Papalini, L.~Russo, and D.~Seminara, ``{Asymptotics of Weil-Petersson volumes and two-dimensional quantum gravities},'' \href{http://arxiv.org/abs/2402.07276}{{\ttfamily arXiv:2402.07276 [hep-th]}}.

\bibitem{Blommaert:2019wfy}
A.~Blommaert, T.~G. Mertens, and H.~Verschelde, ``{Eigenbranes in Jackiw-Teitelboim gravity},'' \href{http://dx.doi.org/10.1007/JHEP02(2021)168}{{\em JHEP} {\bfseries 02} (2021) 168}, \href{http://arxiv.org/abs/1911.11603}{{\ttfamily arXiv:1911.11603 [hep-th]}}.

\bibitem{Marolf:2020xie}
D.~Marolf and H.~Maxfield, ``{Transcending the ensemble: baby universes, spacetime wormholes, and the order and disorder of black hole information},'' \href{http://dx.doi.org/10.1007/JHEP08(2020)044}{{\em JHEP} {\bfseries 08} (2020) 044}, \href{http://arxiv.org/abs/2002.08950}{{\ttfamily arXiv:2002.08950 [hep-th]}}.

\bibitem{Blommaert:2021fob}
A.~Blommaert, L.~V. Iliesiu, and J.~Kruthoff, ``{Gravity factorized},'' \href{http://dx.doi.org/10.1007/JHEP09(2022)080}{{\em JHEP} {\bfseries 09} (2022) 080}, \href{http://arxiv.org/abs/2111.07863}{{\ttfamily arXiv:2111.07863 [hep-th]}}.

\bibitem{Saad:2019lba}
P.~Saad, S.~H. Shenker, and D.~Stanford, ``{JT gravity as a matrix integral},'' \href{http://arxiv.org/abs/1903.11115}{{\ttfamily arXiv:1903.11115 [hep-th]}}.

\bibitem{Dong:2018cuv}
X.~Dong, E.~Silverstein, and G.~Torroba, ``{De Sitter Holography and Entanglement Entropy},'' \href{http://dx.doi.org/10.1007/JHEP07(2018)050}{{\em JHEP} {\bfseries 07} (2018) 050}, \href{http://arxiv.org/abs/1804.08623}{{\ttfamily arXiv:1804.08623 [hep-th]}}.

\bibitem{Mertens:2019tcm}
T.~G. Mertens and G.~J. Turiaci, ``{Defects in Jackiw-Teitelboim Quantum Gravity},'' \href{http://dx.doi.org/10.1007/JHEP08(2019)127}{{\em JHEP} {\bfseries 08} (2019) 127}, \href{http://arxiv.org/abs/1904.05228}{{\ttfamily arXiv:1904.05228 [hep-th]}}.

\bibitem{Yang:2018gdb}
Z.~Yang, ``{The Quantum Gravity Dynamics of Near Extremal Black Holes},'' \href{http://dx.doi.org/10.1007/JHEP05(2019)205}{{\em JHEP} {\bfseries 05} (2019) 205}, \href{http://arxiv.org/abs/1809.08647}{{\ttfamily arXiv:1809.08647 [hep-th]}}.

\bibitem{Blommaert:2023vbz}
A.~Blommaert, J.~Kruthoff, and S.~Yao, ``{The power of Lorentzian wormholes},'' \href{http://arxiv.org/abs/2302.01360}{{\ttfamily arXiv:2302.01360 [hep-th]}}.

\bibitem{Lam:2018pvp}
H.~T. Lam, T.~G. Mertens, G.~J. Turiaci, and H.~Verlinde, ``{Shockwave S-matrix from Schwarzian Quantum Mechanics},'' \href{http://dx.doi.org/10.1007/JHEP11(2018)182}{{\em JHEP} {\bfseries 11} (2018) 182}, \href{http://arxiv.org/abs/1804.09834}{{\ttfamily arXiv:1804.09834 [hep-th]}}.

\bibitem{Maldacena:2015waa}
J.~Maldacena, S.~H. Shenker, and D.~Stanford, ``{A bound on chaos},'' \href{http://dx.doi.org/10.1007/JHEP08(2016)106}{{\em JHEP} {\bfseries 08} (2016) 106}, \href{http://arxiv.org/abs/1503.01409}{{\ttfamily arXiv:1503.01409 [hep-th]}}.

\bibitem{HVerlindetalk}
H.~Verlinde, ``{Duality between SYK and 2+1 dimensional de Sitter}.'' Talks given at the QGQC5 conference, UC Davis, August 2019, the Franqui Symposium, Brussels, November 2019, at ‘Quantum Gravity on Southern Cone’, Argentina, December 2019, and ‘SYK models and Gauge Theory’ workshop at Weizmann Institute, December 2019.

\bibitem{Mertens:2020hbs}
T.~G. Mertens and G.~J. Turiaci, ``{Liouville quantum gravity -- holography, JT and matrices},'' \href{http://dx.doi.org/10.1007/JHEP01(2021)073}{{\em JHEP} {\bfseries 01} (2021) 073}, \href{http://arxiv.org/abs/2006.07072}{{\ttfamily arXiv:2006.07072 [hep-th]}}.

\bibitem{Fan:2021bwt}
Y.~Fan and T.~G. Mertens, ``{From quantum groups to Liouville and dilaton quantum gravity},'' \href{http://dx.doi.org/10.1007/JHEP05(2022)092}{{\em JHEP} {\bfseries 05} (2022) 092}, \href{http://arxiv.org/abs/2109.07770}{{\ttfamily arXiv:2109.07770 [hep-th]}}.

\bibitem{Zamolodchikov:2001ah}
A.~B. Zamolodchikov and A.~B. Zamolodchikov, ``{Liouville field theory on a pseudosphere},'' \href{http://arxiv.org/abs/hep-th/0101152}{{\ttfamily arXiv:hep-th/0101152}}.

\bibitem{Martinec:2003ka}
E.~J. Martinec, ``{The Annular report on noncritical string theory},'' \href{http://arxiv.org/abs/hep-th/0305148}{{\ttfamily arXiv:hep-th/0305148}}.

\bibitem{Collier:2023cyw}
S.~Collier, L.~Eberhardt, B.~M\"uhlmann, and V.~A. Rodriguez, ``{The Virasoro Minimal String},'' \href{http://arxiv.org/abs/2309.10846}{{\ttfamily arXiv:2309.10846 [hep-th]}}.

\bibitem{Blommaert:2025avl}
A.~Blommaert, A.~Levine, T.~G. Mertens, J.~Papalini, and K.~Parmentier, ``{Wormholes, branes and finite matrices in sine dilaton gravity},'' \href{http://arxiv.org/abs/2501.17091}{{\ttfamily arXiv:2501.17091 [hep-th]}}.

\bibitem{knizhnik1988fractal}
V.~G. Knizhnik, A.~M. Polyakov, and A.~B. Zamolodchikov, ``Fractal structure of 2d—quantum gravity,'' {\em Modern Physics Letters A} {\bfseries 3} no.~08, (1988) 819--826.

\bibitem{Turiaci:2020fjj}
G.~J. Turiaci, M.~Usatyuk, and W.~W. Weng, ``{2D dilaton-gravity, deformations of the minimal string, and matrix models},'' \href{http://dx.doi.org/10.1088/1361-6382/ac25df}{{\em Class. Quant. Grav.} {\bfseries 38} no.~20, (2021) 204001}, \href{http://arxiv.org/abs/2011.06038}{{\ttfamily arXiv:2011.06038 [hep-th]}}.

\bibitem{Mertens:2022aou}
T.~G. Mertens, ``{Quantum exponentials for the modular double and applications in gravity models},'' \href{http://dx.doi.org/10.1007/JHEP09(2023)106}{{\em JHEP} {\bfseries 09} (2023) 106}, \href{http://arxiv.org/abs/2212.07696}{{\ttfamily arXiv:2212.07696 [hep-th]}}.

\bibitem{toap}
A.~Blommaert, T.~G. Mertens, and J.~Papalini, ``{Work in progress},''.

\bibitem{Gao:2021uro}
P.~Gao, D.~L. Jafferis, and D.~K. Kolchmeyer, ``{An effective matrix model for dynamical end of the world branes in Jackiw-Teitelboim gravity},'' \href{http://dx.doi.org/10.1007/JHEP01(2022)038}{{\em JHEP} {\bfseries 01} (2022) 038}, \href{http://arxiv.org/abs/2104.01184}{{\ttfamily arXiv:2104.01184 [hep-th]}}.

\bibitem{Blommaert:2022ucs}
A.~Blommaert, L.~V. Iliesiu, and J.~Kruthoff, ``{Alpha states demystified \textemdash{} towards microscopic models of AdS$_{2}$ holography},'' \href{http://dx.doi.org/10.1007/JHEP08(2022)071}{{\em JHEP} {\bfseries 08} (2022) 071}, \href{http://arxiv.org/abs/2203.07384}{{\ttfamily arXiv:2203.07384 [hep-th]}}.

\bibitem{Okuyama:2021eju}
K.~Okuyama and K.~Sakai, ``{FZZT branes in JT gravity and topological gravity},'' \href{http://dx.doi.org/10.1007/JHEP09(2021)191}{{\em JHEP} {\bfseries 09} (2021) 191}, \href{http://arxiv.org/abs/2108.03876}{{\ttfamily arXiv:2108.03876 [hep-th]}}.

\bibitem{Saad:2021uzi}
P.~Saad, S.~Shenker, and S.~Yao, ``{Comments on wormholes and factorization},'' \href{http://arxiv.org/abs/2107.13130}{{\ttfamily arXiv:2107.13130 [hep-th]}}.

\bibitem{Okuyama:2023byh}
K.~Okuyama, ``{End of the world brane in double scaled SYK},'' \href{http://dx.doi.org/10.1007/JHEP08(2023)053}{{\em JHEP} {\bfseries 08} (2023) 053}, \href{http://arxiv.org/abs/2305.12674}{{\ttfamily arXiv:2305.12674 [hep-th]}}.

\bibitem{Belaey:2023jtr}
A.~Belaey, F.~Mariani, and T.~G. Mertens, ``{Branes in JT (super)gravity from group theory},'' \href{http://dx.doi.org/10.1007/JHEP02(2024)058}{{\em JHEP} {\bfseries 02} (2024) 058}, \href{http://arxiv.org/abs/2310.04245}{{\ttfamily arXiv:2310.04245 [hep-th]}}.

\bibitem{Anninos:2017hhn}
D.~Anninos and D.~M. Hofman, ``{Infrared Realization of dS$_2$ in AdS$_2$},'' \href{http://dx.doi.org/10.1088/1361-6382/aab143}{{\em Class. Quant. Grav.} {\bfseries 35} no.~8, (2018) 085003}, \href{http://arxiv.org/abs/1703.04622}{{\ttfamily arXiv:1703.04622 [hep-th]}}.

\bibitem{Anninos:2020cwo}
D.~Anninos and D.~A. Galante, ``{Constructing AdS$_{2}$ flow geometries},'' \href{http://dx.doi.org/10.1007/JHEP02(2021)045}{{\em JHEP} {\bfseries 02} (2021) 045}, \href{http://arxiv.org/abs/2011.01944}{{\ttfamily arXiv:2011.01944 [hep-th]}}.

\bibitem{Kleinert:788154}
H.~Kleinert, {\em {Path integrals in quantum mechanics, statistics, polymer physics, and financial markets; 3rd ed.}}
\newblock World Scientific, River Edge, NJ, 2004.
\newblock Based on a Course on Path Integrals, Freie Univ. Berlin, 1989/1990.

\end{thebibliography}\endgroup

\end{document}